\documentclass[draft]{agujournal2019}
\usepackage[T1]{fontenc}
\usepackage[latin9]{inputenc}
\usepackage{url}
\usepackage{amsmath}
\usepackage{amssymb}
\usepackage{graphicx}
\usepackage[authoryear]{natbib}

\makeatletter

\providecommand{\tabularnewline}{\\}

\usepackage{url} 
\usepackage[inline]{trackchanges} 
\usepackage{soul,slashbox}
\draftfalse

\journalname{Journal of Advances in Modeling Earth Systems (JAMES)}

\makeatletter
\newcommand{\biggg}{\bBigg@\thr@@}
\makeatother
\makeatother

\begin{document}
\noindent \title{Adaptively Implicit Advection for Atmospheric Flows}

\authors{
Hilary Weller\affil{1},
Christian K\"uhnlein\affil{2}, and
Piotr K. Smolarkiewicz\affil{3}
}
\affiliation{1}{Meteorology, University of Reading, RG6 6BB,UK}
\affiliation{2}{ECMWF, 53175 Bonn, Germany}
\affiliation{3}{NSF National Center for Atmospheric Research, Boulder, Colorado 80301, USA}
\correspondingauthor{Hilary Weller}{h.weller@reading.ac.uk}
\begin{keypoints}
\item Implicit time stepping for advection enables huge Courant nubmers.
\item Implicit time stepping for advection is cheap as it is applied as an explicit correction on first-order implicit.
\item Implicit time stepping enables monotonicity, conservation and unconditional stability.
\end{keypoints}
\begin{abstract}
Implicit time-stepping for advection is applied locally in space and
time where Courant numbers are large, but standard explicit time-stepping
is used for the remaining solution which is typically the majority.
This adaptively implicit advection scheme facilitates efficient and
robust integrations with long time-steps while having negligible impact
on the overall accuracy, and achieving monotonicity and local conservation
on general meshes. A novel and important aspect for the efficiency
of the approach is that only one linear solver iteration is needed
for each advection solve. 

The implementation in this paper uses a second-order Runge-Kutta implicit/explicit
time-stepping in combination with a second/third-order finite volume
spatial discretisation. We demonstrate the adaptively implicit advection
in the context of deformational flow advection on the sphere and a
fully compressible model for atmospheric flows. Tracers are advected
over the poles of latitude-longitude grids with very large Courant
numbers and through hexagonal and cubed-sphere meshes with the same
algorithm. Buoyant flow simulations with strong local updrafts also
benefit from adaptively implicit advection. Stably stratified flow
simulations require a stable combination of implicit treatment of
gravity and acoustic waves as well as advection in order to achieve
long stable time-steps.
\end{abstract}

\section*{Plain Language Summary}

Weather and climate prediction models take small time-steps in order
to make predictions about the future, starting from estimates of current
conditions. The smaller the time-steps are, the more of them have
to be taken to make a prediction for a given time in the future. The
more time-steps that have to be taken, the more expensive the prediction
is. If the time-steps are too big, models can not only lose accuracy,
they can become unstable -- inaccuracies can become so large that
wild oscillations are generated and the model crashes. These instabilities
are often caused by the transport (advection) of constituents of the
atmosphere by the wind. This paper describes a method -- adaptive
implicit advection -- for calculating atmospheric transport using
longer time-steps while maintaining stability. We show that this can
be achieved with minimal additional cost, and accuracy is only lost
locally, where the time-step is large relative to the flow speed and
model grid size.

\section{Introduction}

Weather and climate models can have severe time-step restrictions
due to advection \citep[e.g.][]{CD91,DEE+12,KDK+19,MBS+19,GKC13}.
Historically these were most problematic near the poles of latitude-longitude
grids, and could be circumvented using techniques such as polar filtering
\citep{CD91} and semi-Lagrangian advection \citep{Rob82,PS84,PL91,Hort02,DCM+05}.
State-of-the-art atmospheric models have moved away from full latitude-longitude
grids \citep[e.g.][]{DEE+12,GKC13,KDK+19,MBS+19} but models that
use explicit Eulerian schemes for advection still have significant
time-step restrictions where high velocities occur and coincide with
small spatial grid increments, e.g. in storms, convection, jets, and
in interaction with the underlying terrain.

Semi-Lagrangian is the most common approach to address time step limitations.
However, for the typical interpolating semi-Lagrangian schemes, this
occurs at the expense of conservation. Semi-Lagrangian schemes are
most straightforward to implement on structured grids, such as the
latitude-longitude grid. Nevertheless, efficient advective-form (i.e.,
non-conservative) algorithms that allow very long time steps have
been developed for the quasi-uniform octahedral grid in ECMWF\textquoteright s
operational IFS \citep{MWD+16,DV22}

Conservative, cell-integrated, or flux-form semi-Lagrangian (FFSL)
schemes \citep[e.g.][]{LR96,ZWS04,Lau07} entail mapping the old time-step
solution onto non-overlapping departure cells (rather than interpolating
onto departure points). This carries additional cost and complexity
but enables long time-steps without sacrificing conservation. FFSL
entails calculating fluxes of the quantity advected through faces
each time-step. It can be made equivalent to cell-integrated semi-Lagrangian
for small Courant numbers \citep{LEM11}. For large Courant numbers,
the swept volumes pass through multiple cells and the contribution
from each cell is accounted for, leading to a cost that scales with
Courant number. On structured grids this cost can be small \citep{LLM95,LLM96,LR96}
and accuracy is retained for large time-steps \citep{CWPS17}. In
practice, FFSL schemes are often used with small Courant numbers \citep{LR05b,PL07,Miu07b,SM10,HLM11,LUJ+14}
but the technique is being actively investigated by the UK Met Office
for use with large Courant numbers on a cubed-sphere grid (James Kent,
pers. comm. 2023).

Implicit time-stepping for advection has been used in computational
fluid dynamics for decades \citep[e.g.][]{YH87} and is gaining popularity
in atmosphere and ocean models \citep[e.g.][]{BSF+11,JKW11,Shch15,WS20,WWKS23,LD24}.
There are some commonly held beliefs within atmospheric modelling
about implicit time-stepping for advection that warrant further investigation:
\begin{enumerate}
\item Implicit time-stepping for advection is not worth the significant
cost of the additional linear equation solves. In particular, the
additional linear equation solves will entail a cost proportional
to the time-step, implying that sub-steps for advection, or just smaller
time-steps, would be more cost effective.
\item Additional implicit time stepping schemes will reduce parallel scaling
performance.
\item Implicit time-stepping entails severe advection phase errors.
\end{enumerate}
There is also a mathematical order barrier which states that no implicit
method exists which is monotonic for all time-steps and has order
greater than one in time \citep{GST01}. Therefore, to create useful
implicit method for advection, it is necessary to either drop the
need for strict monotonicity \citep{YH87} or downgrade to first-order
accuracy where Courant numbers are large, which can be done using
implicit extensions of flux-corrected transport \citep[FCT,][]{MB17,WWKS23}.

This paper builds on \citealt{WWKS23} (WWKS23) who created an adaptive
implicit version of MPDATA \citep{SM98,SS05b}, which enabled arbitrarily
large Courant numbers and was tested on deformational flow on a variety
of meshes of the sphere, including with Courant numbers up to 70 over
the pole of a rotated latitude-longitude grid. WWKS23 used two complementary
types of adaptivity: (i) implicit advection only where the Courant
number was above one, (ii) the spatial and temporal order of accuracy
gradually reduced to one for Courant numbers above one. Accuracy greater
than first-order was achieved using explicit corrections, so the linear
equation systems were sparse and diagonally dominant. There were shortcomings
of this work:
\begin{itemize}
\item The stabilisation necessary for large Courant numbers was stronger
than predicted by theory.
\item Number of matrix equation solver iterations scaled close to linearly
with the domain mean Courant number.
\item Only linear advection test cases in two spatial dimensions were presented.
\end{itemize}
In this paper we present an advection scheme with the following properties,
which are assumed sufficient when the Courant number is only above
one in small regions of the atmosphere:
\begin{itemize}
\item Accuracy and cost comparable with explicit time-stepping where the
Courant number is less than one.
\item Stable and at least first-order accurate for Courant numbers larger
than one.
\item Monotone for all Courant numbers (no spurious extrema generated).
\item Suitable for quasi-uniform meshes of the sphere, unstructured meshes
and meshes with local adaptivity.
\item Exactly locally conservative.
\item Multi-tracer efficient.
\item Good parallel scalability.
\item Coupled with a solution of the Navier-Stokes equations with implicit
treatment of acoustic and gravity waves.
\end{itemize}
This is achieved by using method-of-lines (Runge-Kutta) adaptively
implicit time-stepping with a quasi-cubic upwind spatial discretisation
of advection terms coupled to a co-located (A-grid), second-order,
finite volume Navier-Stokes solver. We use a segregated, un-split
approach between advection and other terms; the advection terms of
the momentum and potential temperature equations are treated implicitly
(where needed) while terms involving acoustic and gravity wave propagation
are held fixed. Next, the advection terms are held fixed while acoustic
and gravity waves are treated implicitly for the solution of the Helmholtz
(pressure) equation. Outer iterations are necessary for convergence.
The method-of-lines advection scheme is described in section \ref{subsec:implicitAdvection}.
The linear stability of the advection scheme with implicit treatment
of gravity waves is analysed in section \ref{subsec:NA}. The coupling
of the adaptively implicit advection with the Navier-Stokes equations
is described in section \ref{subsec:NSsolution} and \ref{appx:NSsolution}.
All numerical schemes are implemented and tested in the OpenFOAM 11
C++ toolbox \citep{OpenFOAM24}. Deformational flow linear advection
test cases on the sphere are presented in section \ref{subsec:advectionResults},
demonstrating that the method-of-lines scheme has slightly improved
performance to the implicit MPDATA of WWKS23. Section \ref{subsec:bouyantFlowResults}
shows results of buoyancy driven flows and section \ref{subsec:bouyantFlowResults}
shows results of stably stratified flow over orography. These tests
show the benefit of implicit treatment of gravity waves as well as
acoustic waves and advection. Conclusions are drawn in section \ref{sec:concs}.

\section{Numerical Modelling}

The discretisation for advection and for other terms in the Navier-Stokes
equations is split in space and time so that the PDEs are converted
to ODEs using the spatial discretisation described in sections \ref{subsec:OpenFOAMspace}
and \ref{subsec:qCubicUpwind}. The ODEs are then solved with an implicit/explicit
(IMEX) Runge-Kutta scheme with adaptively implicit advection (section
\ref{subsec:implicitAdvection}). The application of this scheme to
the linear advection equation is described in \ref{subsec:ImExRK}
and to the Navier-Stokes equations in \ref{subsec:NSsolution} and
\ref{appx:NSsolution}. The presented time integration technique should
be applicable with any spatial discretisation.

The stability of the adaptive implicit advection coupled with implicit
treatment of gravity waves is analysed in section \ref{subsec:NA}.
This analysis is used to define (i) the parameter $\alpha$ that controls
off-centring between old and new time levels, (ii) $\beta$ that controls
implicit advection and (iii) $\gamma$ that controls the higher-order
spatial correction. A summary of the notation is given in \ref{appx:notation}.

\subsection{Overview of OpenFOAM Spatial Discretisation\label{subsec:OpenFOAMspace}}

The spatial discretisation is composed from a selection of OpenFOAM
version 11 finite volume operators for arbitrary meshes \citep{OpenFOAM24}
and a simple quasi-cubic upwind advection scheme described in section
\ref{subsec:qCubicUpwind}. 

\noindent 
\begin{figure}
\noindent \begin{centering}
\includegraphics[width=0.4\textwidth]{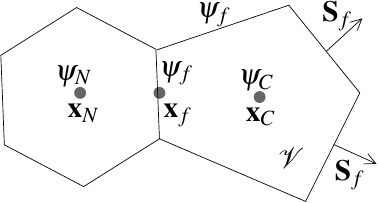}
\par\end{centering}
\caption{Cell $C$, with neighbour $N$, straddling face $f$ in an arbitrary
mesh. Cells have centres $\mathbf{x}_{C}$ and $\mathbf{x}_{N}$ and
face $f$ has centre $\mathbf{x}_{f}$. Cell centre quantity $\psi$
has value $\psi_{f}$ at the face. Cells have volume $\mathcal{V}$.
Faces have area vectors $\mathbf{S}_{f}$ normal to the face with
magnitude equal to the face area.\label{fig:twoCells}}
\end{figure}
This section describes the spatial discretisation of a generic variable
$\psi$ for terms other than advection. Figure \ref{fig:twoCells}
shows a cell with value $\psi_{C}$ at cell $C$ and $\psi_{N}$ at
a neighbour cell $N$, with face $f$ in between. Cells have volume
$\mathcal{V}$ and faces have outward pointing area vectors $\mathbf{S}_{f}$
which are normal to the face with magnitude equal to the face area.

Using Gauss's theorem, cell gradients of $\psi$ in volume $\mathcal{V}$
are calculated as 
\begin{equation}
\nabla\psi=\int_{\mathcal{V}}\nabla\psi\ \text{d}\mathcal{V}=\frac{1}{\mathcal{V}}\int_{S}\psi\text{d}\mathbf{S}\approx\frac{1}{\mathcal{V}}\sum_{\text{faces}}\left\{ \psi\right\} _{f}\mathbf{S}_{f},\label{eq:GaussGrad}
\end{equation}
where operator $\left\{ \right\} _{f}$ means linear interpolation
from cells to faces
\begin{equation}
\left\{ \psi\right\} _{f}=\lambda\psi_{C}+\left(1-\lambda\right)\psi_{N},\label{eq:linInterp}
\end{equation}
where $\lambda$ is the linear interpolation weight from points $\mathbf{x}_{C}$
and $\mathbf{x}_{N}$ (the cell centres) to $\mathbf{x}_{f}$ (the
face centre)
\begin{equation}
\lambda=\frac{|\mathbf{x}_{N}-\mathbf{x}_{f}|}{|\mathbf{x}_{N}-\mathbf{x}_{C}|}.
\end{equation}
Gradients are calculated at faces by linearly interpolating cell gradients
and correcting the component normal to the face using the compact,
two-point difference
\begin{equation}
\nabla_{f}\psi=\left\{ \nabla\psi\right\} _{f}+\left(\psi_{N}-\psi_{C}\right)\frac{\mathbf{x}_{N}-\mathbf{x}_{C}}{|\mathbf{x}_{N}-\mathbf{x}_{C}|^{2}}-\left\{ \nabla\psi\right\} _{f}\cdot\frac{\mathbf{x}_{N}-\mathbf{x}_{C}}{|\mathbf{x}_{N}-\mathbf{x}_{C}|}.\label{eq:sfGrad}
\end{equation}
Scalar valued, surface normal gradients, $\nabla_{S}\psi=\mathbf{S}_{f}\cdot\nabla\psi$,
use a non-orthogonal correction of the two-point difference
\begin{equation}
\nabla_{S}\psi=\frac{\psi_{N}-\psi_{C}}{|\mathbf{d}_{f}|}|\mathbf{S}_{f}|+\left\{ \nabla\psi\right\} _{f}\cdot\mathbf{S}_{f}-\left\{ \nabla\psi\right\} _{f}\cdot\mathbf{d}_{f},\label{eq:snGrad}
\end{equation}
where $\mathbf{d}_{f}=\mathbf{x}_{N}-\mathbf{x}_{C}$. The first term
of (\ref{eq:snGrad}) is the two-point difference, $\nabla_{2}\psi$,
and the second two terms are the non-orthogonal correction, $\nabla_{\text{noc}}\psi$.
The two point difference can be treated implicitly and the non-orthogonal
correction is always treated explicitly. Note that $\nabla$ is used
as a gradient operator even when it only refers to a component of
the gradient.

There are two relevant spatial averaging operators denoted by the
symbol $\left\{ \right\} _{C}$ which are the reconstruction of cell
centre vectors, $\mathbf{a}_{C}$, from face normal vector $A_{S}=\mathbf{a}_{f}\cdot\mathbf{S}_{f}$,
and averaging of face scalar values, $\Psi_{f}$, to cell centres,
defined as
\begin{eqnarray}
\mathbf{a}_{C}=\left\{ A_{S}\right\} _{C} & = & \left(\sum_{\text{faces}}\mathbf{S}_{f}\mathbf{S}_{f}^{T}\right)^{-1}\sum_{\text{faces}}\mathbf{S}_{f}A_{S},\\
\left\{ \psi_{f}\right\} _{C} & = & \sum_{\text{faces}}|\mathbf{S}_{f}|\psi_{f}\bigg/\sum_{\text{faces}}|\mathbf{S}_{f}|,
\end{eqnarray}
where $\sum_{\text{faces}}\mathbf{S}_{f}\mathbf{S}_{f}^{T}$ is a
$3\times3$ tensor for each cell.

\subsection{Quasi-Cubic Upwind Advection\label{subsec:qCubicUpwind}}

Divergence is calculated in cells with volume $\mathcal{V}$ using
Gauss's theorem
\begin{equation}
\nabla\cdot\rho\mathbf{u}\psi=\frac{1}{\mathcal{V}}\int_{\mathcal{V}}\nabla\cdot\rho\mathbf{u}\psi\ \text{d}\mathcal{V}=\frac{1}{\mathcal{V}}\int_{S}\psi\rho\mathbf{u}\cdot\text{d}\mathbf{S}\approx\frac{1}{\mathcal{V}}\sum_{\text{faces}}\phi\psi_{f},\label{eq:GaussDiv}
\end{equation}
where $\rho$ is the fluid density, $\mathbf{u}$ is the velocity
vector and $\phi$ is the mass flux through face $f$ given as
\begin{equation}
\phi=\rho_{f}\mathbf{u}_{f}\cdot\mathbf{S}_{f}.\label{eq:definePhi}
\end{equation}
Quasi third-order accurate advection is implemented as a correction
to the first-order upwind scheme defined as a blend, $b$, of two
gradients
\begin{equation}
\psi_{f}=\underbrace{\psi_{u}}_{\text{1st order upwind}}+\gamma\underbrace{\left(\mathbf{x}_{f}-\mathbf{x}_{u}\right)\cdot\left(b\nabla_{u}\psi+\left(1-b\right)\nabla_{f}\psi\right)}_{\sim\text{3rd order correction (HOC)}},\label{eq:cubicf}
\end{equation}
where 
\begin{eqnarray*}
\text{the upwind cell, }u & = & \begin{cases}
N & \text{if }\phi>0\\
C & \text{otherwise},
\end{cases}\\
\text{the downwind cell, }d & = & \begin{cases}
N & \text{if }\phi\le0\\
C & \text{otherwise.}
\end{cases}
\end{eqnarray*}
The parameter $\gamma\in(0,1]$ is the higher-order flux limiter (defined
in section \ref{subsec:NA}). The blend $b$ is defined to attain
third-order accuracy in representing $\partial\psi/\partial x$ on
a one-dimensional, uniform grid, indexed by $j$ ($\psi_{j}=\psi_{u}$
and $\psi_{j+1}=\psi_{d}$ for $u>0$)
\begin{equation}
\frac{\partial\psi}{\partial x}_{j}\approx\frac{2\psi_{j+1}+3\psi_{j}-6\psi_{j-1}+\psi_{j-2}}{6\Delta x}.
\end{equation}
We express this as a difference between face values as
\begin{eqnarray}
\frac{\partial\psi}{\partial x}_{j} & = & \frac{\psi_{j+\frac{1}{2}}-\psi_{j-\frac{1}{2}}}{\Delta x}\implies\psi_{j+\frac{1}{2}}=\frac{2\psi_{j+1}+5\psi_{j}-\psi_{j-1}}{6}.
\end{eqnarray}
In (\ref{eq:cubicf}) the 3rd-order correction (high-order correction,
HOC) is expressed as a linear combination of gradients. On a one-dimensional,
uniform grid, these gradients are given as
\[
\nabla_{u}\psi=\frac{\psi_{j+1}-\psi_{j-1}}{2\Delta x},\ \nabla_{f}\psi=\frac{\psi_{j+1}-\psi_{j}}{\Delta x}\text{ and }\mathbf{x}_{f}-\mathbf{x}_{u}=\frac{\Delta x}{2}.
\]
Substituting these into (\ref{eq:cubicf}) gives 
\begin{eqnarray}
\frac{2\psi_{j+1}+5\psi_{j}-\psi_{j-1}}{6} & = & \psi_{j}+\frac{\Delta x}{2}\left(b\frac{\psi_{j+1}-\psi_{j-1}}{2\Delta x}+\left(1-b\right)\frac{\psi_{j+1}-\psi_{j}}{\Delta x}\right),\nonumber \\
\implies b & = & \frac{2}{3}.\label{eq:cublicBlendb}
\end{eqnarray}
$b=\frac{2}{3}$ is then used for arbitrary two- and three-dimensional
meshes in (\ref{eq:cubicf}), which provides third-order accuracy
for $\gamma=1$ in highly idealised settings when the meshes are uniform,
as verified in \ref{appx:cubicConvergence}.

\subsection{Adaptive Implicit Time Stepping for Advection-Diffusion\label{subsec:implicitAdvection}}

\subsubsection{Adaptively Implicit Runge-Kutta Scheme \label{subsec:ImExRK}}

The adaptively implicit time-stepping provides a solution of the linear
advection-diffusion equation
\begin{equation}
\frac{\partial\rho\psi}{\partial t}+\nabla\cdot\rho\mathbf{u}\psi-\nabla\cdot\left(\rho K\nabla\psi\right)=S_{\psi},\label{eq:advectDiffuse}
\end{equation}
for transported quantity $\psi$, where $\rho$ is the density, $\mathbf{u}$
is the velocity vector and $K$ the diffusion coefficient, while $S_{\psi}$
subsumes source terms. The mass flux and the density satisfy the mass
continuity equation discretely which is given the  shorthand
\begin{equation}
\frac{\partial\rho}{\partial t}=-\frac{1}{\mathcal{V}}\sum_{\text{faces}}\phi_{f}=-\nabla\cdot\phi.
\end{equation}
For $\psi=\mathbf{u}$, the corresponding source $S_{\mathbf{u}}$
contains the pressure gradient and gravity terms.

In the description of the advection of $\psi$, the advecting velocity
always appears with the density and it is always needed as a flux
on faces. We therefore write the semi-discretised equations using
the mass flux $\phi$ defined in (\ref{eq:definePhi}). To reduce
the complexity of the presentation, $\phi$ can also symolise a vector
mass flux under the divergence operator. The time-stepping is framed
as a Runge-Kutta IMEX (implicit/explicit) scheme and is based on trapezoidal
implicit with off-centring. This is combined with an explicit correction
of high-order (HO) terms (HOC being the high-order correction)
\begin{eqnarray}
\frac{\rho^{k}\psi^{k}-\rho^{(n)}\psi^{(n)}}{\Delta t}= & - & \nabla\cdot\left[\left(1-\alpha\right)\phi^{(n)}\psi_{\text{HO}}^{(n)}\right]-\nabla\cdot\text{\ensuremath{\left[\alpha\left(1-\beta\right)\phi^{k-1}\psi_{\text{HO}}^{k-1}\right]}}\label{eq:RK_ImEx}\\
 & - & \nabla\cdot\left[\alpha\beta\phi^{k-1}\psi_{1\text{st}}^{k}\right]-\nabla\cdot\left[\alpha\beta\gamma\phi^{k-1}\psi_{\text{HOC}}^{k-1}\right]\nonumber \\
 & + & \nabla\cdot\left[\left(1-\alpha\right)\rho_{f}K\nabla_{S}\psi^{(n)}\right]\nonumber \\
 & + & \nabla\cdot\left[\alpha\rho_{f}K\nabla_{2}\psi^{k}\right]+\nabla\cdot\left[\alpha\rho_{f}K\nabla_{\text{noc}}\psi^{k-1}\right]\nonumber \\
 & + & \left(1-\alpha\right)S_{\psi}^{(n)}+\alpha S_{\psi}^{k-1},\nonumber 
\end{eqnarray}
where $\psi^{(n)}$ is at time-step $n$ and outer iterations are
indexed by $k$ with $\psi^{k=0}=\psi^{(n)}$ and 
\begin{equation}
\psi_{\text{HO}}=\psi_{\text{1st}}+\gamma\psi_{\text{HOC}}.
\end{equation}
This again uses the shorthand
\begin{equation}
\frac{1}{\mathcal{V}}\sum_{\text{faces}}\phi\psi=\nabla\cdot\phi\psi.
\end{equation}
The results in section \ref{sec:results} use a maximum $k$ of 2
(implying two outer iterations). In (\ref{eq:RK_ImEx}), the mass
fluxes are updated after $\psi$ so the iteration count is $k-1$
for $\phi$. This is explained in full in \ref{appx:NSsolution}.) 

Consistency with continuity will be addressed in section \ref{subsec:continuityConsistency}.

\subsubsection{Matrix Structure and Solution}

The advection-diffusion equation (\ref{eq:RK_ImEx}) is expressed
as a matrix equation of the form
\begin{equation}
\left(A_{\psi}-H_{\psi}\right)\psi^{k}=R_{\psi}+T_{\psi},\label{eq:AandHpsi}
\end{equation}
where $A_{\psi}$ is a diagonal matrix, $H_{\psi}$ is a matrix with
zeros on the diagonal and $R_{\psi}$ and $T_{\psi}$ are the explicit
right hand sides, consisting of all terms that do not depend on $\psi^{k}$.
In order to calculate $A_{\psi}$ and $H_{\psi}$ we consider the
first-order upwind discretisation of $\nabla\cdot\alpha\beta\phi\psi$
at cell $C$ with neighbours $N$
\begin{eqnarray}
\nabla_{1\text{st}}\cdot\alpha\beta\phi\psi & = & \frac{1}{\mathcal{V}}\sum_{\phi\ge0}\alpha\beta\phi\psi_{C}+\frac{1}{\mathcal{V}}\sum_{\phi<0}\alpha\beta\phi\psi_{N}\\
 & = & \frac{\psi_{C}}{\mathcal{V}}\sum_{\phi\ge0}\alpha\beta\phi-\frac{1}{\mathcal{V}}\sum_{\phi<0}\alpha\beta|\phi|\psi_{N},\nonumber 
\end{eqnarray}
and the two-point discretisation of the diffusion term
\begin{eqnarray*}
\nabla\cdot\text{\ensuremath{\left[\alpha\rho_{f}K\nabla_{2}\psi\right]}} & = & \frac{1}{\mathcal{V}}\sum_{\text{faces}}\alpha\rho_{f}K\frac{\psi_{N}-\psi_{C}}{|\mathbf{d}_{f}|}|\mathbf{S}_{f}|.
\end{eqnarray*}
Therefore $A_{\psi}$ and $H_{\psi}$ take the form
\begin{eqnarray}
A_{\psi} & = & \frac{\rho^{k}}{\Delta t}+\frac{1}{\mathcal{V}_{C}}\sum_{\text{faces}}\alpha\left(\beta\max\left(\phi,0\right)+\rho_{f}K\frac{|\mathbf{S}_{f}|}{|\mathbf{d}_{f}|}\right),\label{eq:Apsi}\\
H_{\psi}\left(C,N\right) & = & \begin{cases}
\frac{\alpha}{\mathcal{V}_{C}}\left(\beta\max\left(-\phi,0\right)+\rho_{f}K\frac{|\mathbf{S}_{f}|}{|\mathbf{d}_{f}|}\right) & \text{for row }C,\text{ column }N\\
0 & \text{otherwise},
\end{cases}\label{eq:Hpsi}
\end{eqnarray}
where $\alpha$, $\beta$ and $\phi$ are on face, $f$, between cells
$C$ and $N$. Note that:
\begin{itemize}
\item for inviscid flow ($K=0$), or if $K$ does not depend on $\psi$,
then $A_{\psi}$ and $H_{\psi}$ do not depend on $\psi$,
\item $A_{\psi}-H_{\psi}$ is diagonally dominant $\forall$ $\alpha\ge0$,
$\beta\ge0$.
\end{itemize}
The right hand side of (\ref{eq:AandHpsi}) contains the previous
time-step, the increments from the first part of the time-steps and
the explicit updates, which are
\begin{eqnarray}
R_{\psi} & = & \underbrace{\frac{\rho^{(n)}\psi^{(n)}}{\Delta t}}_{\text{previous time}}-\underbrace{\nabla\cdot\left(\left(1-\alpha\right)\left[\phi^{(n)}\psi_{\text{HO}}^{(n)}-\rho_{f}K\nabla_{S}\psi^{(n)}\right]\right)+\left(1-\alpha\right)S_{\psi}^{(n)}}_{\text{increments from first part of timestep}}\label{eq:Rpsi}\\
 & - & \underbrace{\nabla\cdot\left(\alpha\left(1-\beta\right)\phi^{k-1}\psi_{\text{HO}}^{k-1}\right)-\nabla\cdot\left(\alpha\beta\gamma\phi^{k-1}\psi_{\text{HOC}}^{k-1}\right)+\nabla\cdot\left(\alpha\rho_{f}K\nabla_{\text{noc}}\psi^{k-1}\right)}_{\text{explicit updates from previous iteration}},\nonumber \\
T_{\psi} & = & \alpha S_{\psi}^{k-1}.\label{eq:Tpsi}
\end{eqnarray}
The reason for writing the matrix in this (OpenFOAM) form is so that
one Jacobi iteration (labelled by j) can be expressed as
\begin{equation}
\psi^{j}=A_{\psi}^{-1}\left(H_{\psi}\psi^{j-1}+R_{\psi}+T_{\psi}\right),
\end{equation}
which will be used in section \ref{subsec:NSsolution}. $T_{\psi}$
is written separately from $R_{\psi}$ because in appendix \ref{subsec:HelmholtzEqn},
$T_{\mathbf{u}}$ includes the pressure gradient term of the momentum
equation when formulating the pressure equation. 

As in WWKS23, ``the resulting linear equation system is solved with
the standard OpenFOAM bi-conjugate gradient solver with a diagonal
incomplete LU preconditioner (DILU)''. A significant difference to
WWKS23 is that no more than one solver iteration is employed per outer
iteration. The proof in WWKS23 that the first-order scheme is monotonic
assumes that the linear system is solved exactly which is not the
case here. However, we will see in section \ref{subsec:slottedCylinder}
that monotonicity is maintained for large Courant numbers with just
one solver iteration. This is straightforward to prove for Jacobi
iterations, but we are not aware of a proof for the DILU solver that
we are using.

\subsubsection{Applying Flux-Corrected Transport (FCT)}

FCT is applied to achieve monotonicity for the monotonic advection
test case (section \ref{subsec:slottedCylinder}). It is not used
in the solution of the Navier-Stokes equations (sections \ref{subsec:bouyantFlowResults},
\ref{subsec:stratifiedFlowResults}), simply because of incompatibilities
between the Navier-Stokes code and the FCT implementation. This is
a situation that needs to be rectified.

Section \ref{subsec:NA} describes the flux limiter $\gamma$, which
is set based on Courant number, to ensure that solutions with large
Courant numbers are smooth. In section \ref{subsec:slottedCylinder}
where FCT is used, $\gamma$ is set to one, so FCT is a different
form of flux limiting to $\gamma$.

The first note of caution in applying FCT to implicit advection is
that we should not expect $\psi_{C}^{(n+1)}$ to be bounded by $\psi_{C}^{(n)}$
and face neighbours $\psi_{N}^{(n)}$, because, with Courant numbers
larger than one, extrema should be able to move beyond neighbouring
cells in one time-step (as discussed by WWKS23). The FCT algorithm
of \citet{Zal79} uses $\psi_{C}^{(n)}$ and $\psi_{N}^{(n)}$ to
bound $\psi_{C}^{(n+1)}$ as well as using the first-order solution
at $n+1$. When using implicit time-stepping, we can only use the
first-order solution at $n+1$ at cell $C$ and neighbours, $N$,
to bound the high-order solution. As a result, this means that applying
FCT to an implicit scheme will be more diffusive than applying it
to an explicit scheme.

The next issue in applying FCT is the requirement of the availability
of a first-order, monotonic solution. This was available in the implicit
version of MPDATA in WWKS23 since MPDATA is expressed as a correction
on the first-order upwind scheme. However, when using RK-IMEX time-stepping,
the first equation solved is (\ref{eq:RK_ImEx}) which gives the high-order
solution. The high-order correction is calculated explicitly, but
the low-order part is solved with the high-order correction as part
of the right hand side. This dramatically improves accuracy in comparison
to first solving a first-order solution and then applying a high-order
correction. This means that, to get the necessary monotonic solution
to use to correct the high-order solution, (\ref{eq:AandHpsi}) must
be solved separately with $\gamma=0$ and $k=1$ to give the monotonic
solution, $\psi^{d}$. Therefore, the monotonic solution, $\psi^{d}$,
is the solution of (\ref{eq:AandHpsi}) but with 
\begin{eqnarray}
R_{\psi} & = & \frac{\rho^{(n)}\psi^{(n)}}{\Delta t}-\nabla\cdot\left(\left(1-\alpha\right)\left[\phi^{(n)}\psi_{1\text{st}}^{(n)}-\rho_{f}K\nabla_{2}\psi^{(n)}\right]\right)+\left(1-\alpha\right)S_{\psi}^{(n)}\\
 & - & \nabla\cdot\left(\alpha\left(1-\beta\right)\phi^{k-1}\psi_{1\text{st}}^{k-1}\right),\nonumber 
\end{eqnarray}
and with $A_{\psi}$ and $H_{\psi}$ the same as for the full solution.
Once $\psi^{d}$, $\psi^{k}$ and the corresponding fluxes are available,
then the FCT algorithm can be applied as described in WWKS23.

\subsection{Numerical Analysis and Stability Parameters\label{subsec:NA}}

Stability analysis is used to set the parameters $\alpha$, $\beta$
and $\gamma$ while optimising accuracy and cost.
\begin{enumerate}
\item The temporal off-centring $\alpha\in[\frac{1}{2},1)$ needs to be
$\frac{1}{2}$ for second-order accuracy in time but can be larger
for improved stability at the expense of temporal accuracy.
\item The implicit/explicit flag $\beta$ is zero where the advection is
calculated explicitly and one where it is implicit.
\item The flux limiter $\gamma\in(0,1]$ limits the high-order spatial correction.
\end{enumerate}
It is essential that $\alpha$, $\beta$ and $\gamma$ are defined
on faces rather than cell centres, and that they are inside the divergence
in (\ref{eq:RK_ImEx}), otherwise local conservation is violated.
However, for the linear numerical analysis, these parameters are assumed
uniform.

The stability of the advection depends on the Courant number, which
can be defined for the finite volume cell of an arbitrary mesh as
\begin{equation}
c=\frac{1}{2}\frac{\Delta t}{\mathcal{V}}\sum_{f}|U_{f}|,\label{eq:c}
\end{equation}
where $U_{f}=\mathbf{u}\cdot\mathbf{S}_{f}$ is the volume fluxes
through each face. In order to set parameters $\alpha$, $\beta$
and $\gamma$ on faces, the Courant number on the face is calculated
as the maximum of the Courant numbers either side of the face
\begin{equation}
c_{f}=\max\left(c_{N},c_{C}\right).\label{eq:cf}
\end{equation}
For stably stratified flow, stability also depends on the Brunt--V\"{a}is\"{a}l\"{a}
frequency defined as
\begin{equation}
N=\sqrt{\max\left(-\frac{\mathbf{g}\cdot\nabla\overline{\theta}}{\overline{\theta}},\ 0\right)},\label{eq:N}
\end{equation}
where $\mathbf{g}=\left(0,0,-g\right)$ is the acceleration due to
gravity and $\overline{\theta}$ is the reference profile potential
temperature (see section \ref{subsec:NSsolution}).

\begin{figure}
\includegraphics[width=1\textwidth]{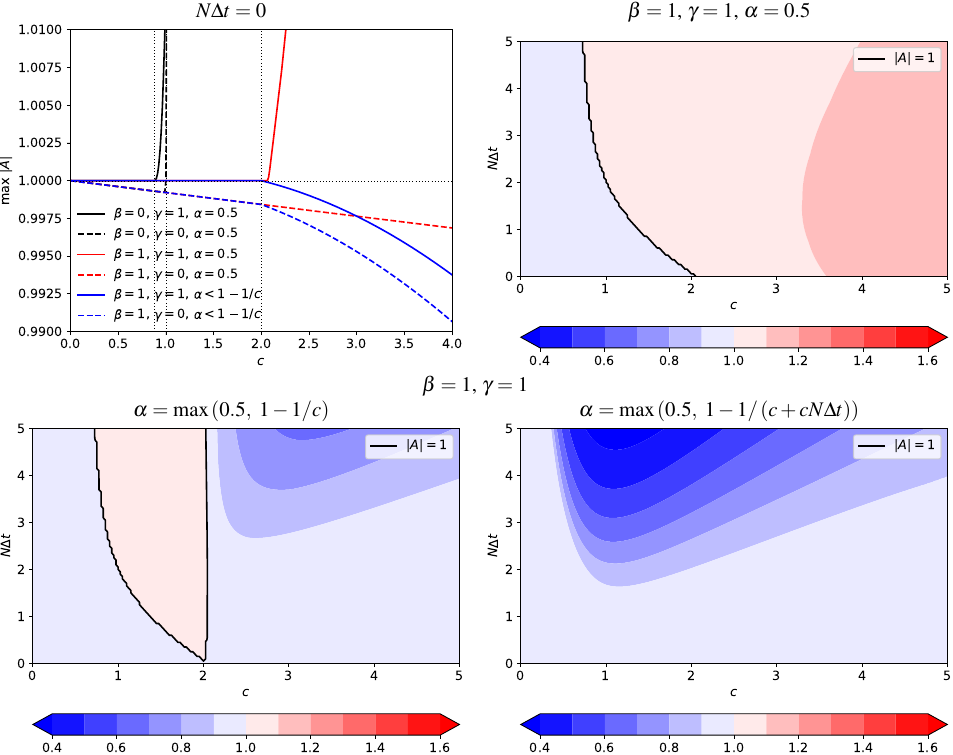}

\caption{Numerical analysis of implicit/explicit cubic upwind solution of (\ref{eq:1dgw}).
In each panel, the maximum of the magnitude of amplification factor
$|\mathcal{A}|$ over all wavenumbers, ${k\Delta x\in\left[0,2\pi\right]}$
is plotted ($\max_{k\Delta x}|\mathcal{A}|$). Top left: $\max_{k\Delta x}|\mathcal{A}|$
as a function of $c$ for $N\Delta t=0$ for schemes with various
combinations of $\alpha$, $\beta$ and $\gamma$. The other panels
show $\max_{k\Delta x}|\mathcal{A}|$ contoured as a function of $c$
and $N\Delta t$ with $|\mathcal{A}|=1$ contoured black, again for
different combinations of $\alpha$, $\beta$ and $\gamma$. The formulation
for $\alpha$ in the bottom right is unconditionally stable so there
is no contour for $|\mathcal{A}|=1$.\label{fig:magA_overkdx}}
\end{figure}

We analyse the simplified, one-dimensional system with advection and
gravity waves
\begin{equation}
\frac{\partial\Psi}{\partial t}+U_{0}\frac{\partial\Psi}{\partial x}=iN\Psi,\label{eq:1dgw}
\end{equation}
where $\Psi=b+iNw$, $i=\sqrt{-1}$, buoyancy $b=g\theta^{\prime}/\overline{\theta}$
and $U_{0}$ is the uniform, constant horizontal velocity. We consider
a uniform grid with spacing $\Delta x$, indexed by $j$ and $c=\Delta tU_{0}/\Delta x$.
The implicit-explicit, 2nd-order Runge-Kutta time-stepping with cubic-upwind
advection is
\begin{eqnarray}
1^{\text{st}}\text{ iteration: }\Psi_{j}^{1} & = & \Psi_{j}^{(n)}-\left(1-\alpha\beta\right)c\left(\Psi_{j}^{(n)}-\Psi_{j-1}^{(n)}\right)-\alpha c\beta\left(\Psi_{j}^{1}-\Psi_{j-1}^{1}\right)\label{eq:RK_1dit1}\\
 & - & \gamma c\left(\Psi_{HOCj+1/2}^{(n)}-\Psi_{HOCj-1/2}^{(n)}\right)\nonumber \\
 & - & iN\Delta t\left(1-\alpha\right)\Psi_{j}^{(n)}-iN\Delta t\alpha\Psi_{j}^{1},\nonumber 
\end{eqnarray}
\begin{eqnarray}
\text{Cubic correction: }\Psi_{HOCj+1/2} & = & \frac{2\Psi_{j+1}-\Psi_{j}-\Psi_{j-1}}{6},\label{eq:1dcubicCorr}
\end{eqnarray}
\begin{eqnarray}
2^{\text{nd}}\text{ iteration: }\Psi_{j}^{(n+1)} & = & \Psi_{j}^{(n)}-\left(1\text{ - }\alpha\right)c\left(\Psi_{j}^{(n)}\text{ - }\Psi_{j-1}^{(n)}\right)-\alpha c\left(1\text{ - }\beta\right)\left(\Psi_{j}^{1}\text{ - }\Psi_{j-1}^{1}\right)\label{eq:RK_1dit2}\\
 & - & \alpha c\beta\left(\Psi_{j}^{(n+1)}-\Psi_{j-1}^{(n+1)}\right)\nonumber \\
 & - & \gamma\left(1-\alpha\right)c\left(\Psi_{HOCj+1/2}^{(n)}-\Psi_{HOCj-1/2}^{(n)}\right)\nonumber \\
 & - & \gamma\alpha c\left(\Psi_{HOCj+1/2}^{1}-\Psi_{HOCj-1/2}^{1}\right)\nonumber \\
 & - & iN\Delta t\left(1-\alpha\right)\Psi_{j}^{(n)}-iN\Delta t\alpha\Psi_{j}^{(n+1)}.\nonumber 
\end{eqnarray}
We consider waves with wavenumber $k$ \footnote{Note that $k$ is used to count outer iterations in other sections
but is used as wavenumber in this section.} in order to perform von-Neumann stability analysis, so that each
wave takes the form
\begin{equation}
\Psi_{j}^{(n)}=\mathcal{A}^{n}e^{ijk\Delta x},\label{eq:AwaveMode}
\end{equation}
where $\mathcal{A}$ is the amplification factor. Substituting (\ref{eq:AwaveMode})
into (\ref{eq:RK_1dit1}-\ref{eq:RK_1dit2}) gives a complicated equation
for $\mathcal{A}$ as a function of $\alpha$, $\beta$, $\gamma$,
$c$, $N\Delta t$ and $k\Delta x$
\begin{equation}
\mathcal{A}=\frac{1-c\left(1-e^{\text{-}ijk\Delta x}\right)\left(1-\alpha+\alpha\left(1\text{ - }\beta\right)\mathcal{A}^{1}\right)-\gamma c\mathcal{A}^{h}\left(1\text{ - }\alpha\text{+ }\alpha\mathcal{A}^{1}\right)-iN\Delta t\left(1\text{ - }\alpha\right)}{1+\alpha c\beta\left(1-e^{-ijk\Delta x}\right)+iN\Delta t\alpha},
\end{equation}
where
\begin{eqnarray*}
\mathcal{A}^{h} & = & \frac{1}{6}\left(2e^{ijk\Delta x}-3+e^{-2ijk\Delta x}\right),\\
\mathcal{A}^{1} & = & \frac{1-\left(1-\alpha\beta\right)c\left(1-e^{-ijk\Delta x}\right)-\gamma c\mathcal{A}^{h}-iN\Delta t\left(1-\alpha\right)}{1+\alpha c\beta\left(1-e^{-ijk\Delta x}\right)+\alpha iN\Delta t}.
\end{eqnarray*}
For stability we require $|\mathcal{A}|\le1\ \forall\ k\Delta x\in\left[0,2\pi\right]$.
Therefore ranges of values of $\alpha$, $\beta$, $\gamma$, $c$
and $N\Delta t$ are considered and the maximum $|\mathcal{A}|$ is
found over 80 equally spaced values of $k\Delta t$. We already have
from WWKS23 and from section \ref{subsec:ImExRK} that for $\gamma=0$
(no high-order in space) we need $\alpha<1-1/c$ for monotonicity.
In order to find when implicit advection is needed ($\beta=1$) in
combination with cubic upwind advection ($\gamma=1$) without interaction
with gravity wave ($N\Delta t=0$), $\max_{k\Delta x\in\left[0,2\pi\right]}\left(|\mathcal{A}|\right)$
is plotted as a function of $c$ for $\beta=0/1$ and $\gamma=0/1$
in the top left panel of figure \ref{fig:magA_overkdx} for, $\alpha=\frac{1}{2}$
and for
\begin{equation}
\alpha=\max\left(\frac{1}{2},\ 1-\frac{1}{c}\right).\label{eq:alpha_c02}
\end{equation}
The explicit scheme ($\beta=0$) with $\alpha=\frac{1}{2}$, is stable
up to $c=1$ for the first-order scheme ($\gamma=0$) and up to $c=0.88$
for cubic upwind $(\gamma=1$). Introducing implicit advection ($\beta=1$)
stabilises first-order advection for all $\alpha\ge\frac{1}{2}$.
However cubic upwind ($\gamma=1$), with $\beta=1$ (implicit) and
with $\alpha=\frac{1}{2}$ (second-order in space) is only stable
for $c\le2$. This is the motivation for eqn (\ref{eq:alpha_c02})
to calculate $\alpha$, which stabilises the implicit cubic upwind
for all Courant numbers. 

\noindent 

The maximum of $|\mathcal{A}|$ over all $k\Delta x\in\left[0,2\pi\right]$
for a range of Courant numbers and a range of $N\Delta t$ is contoured
in figure \ref{fig:magA_overkdx} all using $\beta=1$ (implicit advection)
and $\gamma=1$ (cubic upind advection). The top right showing $|\mathcal{A}|$
for $\alpha=\frac{1}{2}$ confirms the instability for $c>2$ for
$N\Delta t=0$ and shows that this instability occurs for lower Courant
numbers as $N\Delta t$ increases. If we use (\ref{eq:alpha_c02})
to calculate $\alpha$ (as worked for $N\Delta t=0$) then the bottom
left of figure \ref{fig:magA_overkdx} shows that the scheme is now
stable for $c>2$, leaving an instability that occurs for lower Courant
numbers as $N\Delta t$ grows. If instead we use
\begin{equation}
\alpha=\max\left(\frac{1}{2},\ 1-\frac{1}{c+cN\Delta t}\right),\label{eq:alpha_cN}
\end{equation}
(bottom right of figure \ref{fig:magA_overkdx}) then we get stability
for all $c$ and $N\Delta t$ without introducing overly large values
of $\alpha$ for small $c$ for any value of $N\Delta t$.

Based on this analysis we should be able to use $\gamma=1$ for all
$c$. However, we found that $\gamma=1$ can lead to spurious but
stable oscillations when solving the fully compressible Navier-Stokes
equations with large Courant numbers, likely caused by incorrect interactions
between advection and other waves (section \ref{subsec:stratifiedFlowResults}).
Therefore, the value of $\gamma$ is reduced for $c>2$. The final
values of $\alpha$, $\beta$ and $\gamma$ that are used in the solution
of the Navier-Stokes equations are given in table \ref{tab:parameterValues}.
It is possible that a better solution would be to use FCT, but this
has not yet been explored for solving the Navier-Stokes equations
adaptive implicitly.

\noindent 
\begin{table}
\noindent \begin{centering}
\begin{tabular}{cl}
$\alpha:$ & $\max\left(\frac{1}{2},\ 1-\frac{1}{c+cN\Delta t}\right)$\tabularnewline
$\beta:$ & $\begin{cases}
0 & c<0.8\\
1 & c\ge0.8
\end{cases}$\tabularnewline
$\gamma:$ & $\begin{cases}
1 & c<2\\
\frac{4-c}{2} & c\in\left[2,4\right]\\
0 & c>4.
\end{cases}$\tabularnewline
\end{tabular}
\par\end{centering}
\caption{Values of $\alpha$ (temporal off-centring), $\beta$ (implicit switch)
and $\gamma$ (high-order flux limiter) that are used in the numerical
solution of the Navier-Stokes equations. \label{tab:parameterValues}}
\end{table}

\subsection{Solution of the Fully Compressible Navier-Stokes Equations \label{subsec:NSsolution}}

The Navier-Stokes equations are formulated as 
\begin{eqnarray}
\frac{\partial\rho}{\partial t}+\nabla\cdot\rho\mathbf{u} & = & 0,\label{eq:cont}\\
\frac{\partial\rho\theta^{\prime}}{\partial t}+\nabla\cdot\rho\mathbf{u}\theta^{\prime}-\nabla\cdot\left(\rho\kappa\nabla\theta^{\prime}\right) & = & -\rho w\frac{d\overline{\theta}}{dz},\label{eq:theta}\\
\frac{\partial\rho\mathbf{u}}{\partial t}+\nabla\cdot\rho\mathbf{u}\mathbf{u}-\nabla\cdot\left(\rho\nu\nabla\mathbf{u}\right) & = & -\rho\left(\frac{\theta^{\prime}}{\overline{\theta}}\mathbf{g}+c_{p}\theta\nabla\pi^{\prime}\right),\label{eq:mom}
\end{eqnarray}
with the density $\rho$, potential temperature $\theta=T/\pi$, the
velocity vector $\mathbf{u}$, and the Exner pressure defined as $\pi=\left(p/p_{r}\right)^{R/c_{p}}$,
where $R$ is the gas constant of dry air, $c_{p}$ the specific heat
of dry air at constant pressure $p$ and $p_{r}=10^{5}\text{ Pa}$
is a constant references pressure. Furthermore, $\kappa$ is the thermal
diffusivity and $\nu$ is the kinematic viscosity. The primed thermodynamic
variables $\theta^{\prime}=\theta-\overline{\theta}$ and $\pi^{\prime}=\pi-\overline{\pi}$
represent perturbations with respect to reference profiles $\overline{\theta}\left(z\right)$
and $\overline{\pi}\left(z\right)$ in hydrostatic balance given as
\begin{equation}
c_{p}\overline{\theta}\frac{d\overline{\pi}}{dz}=\mathbf{g}\cdot\mathbf{k}=-g.
\end{equation}
The temperature and pressure satisfy the ideal gas law which is written
in terms of perturbation quantities
\begin{equation}
\pi^{\prime}=\left(\frac{\rho R\left(\overline{\theta}+\theta^{\prime}\right)}{p_{r}}\right)^{\frac{R}{c_{v}}}-\overline{\pi},\label{eq:state}
\end{equation}
where $c_{v}=c_{p}-R$ is the specific heat capacity at constant volume.
The formulation in the perturbation variables is beneficial for the
design of the numerical integration procedure, its accuracy and in
the model initialisation.

The advection terms of the momentum and potential temperature equations
are treated implicitly (where needed) while terms involving acoustic
and gravity wave propagation are held fixed, and then the advection
terms are held fixed while acoustic and gravity waves are treated
implicitly for the solution of the Helmholtz (pressure) equation.
Outer iterations are necessary for convergence. 

Details of the solution of the Navier-Stokes equations are provided
in \ref{appx:NSsolution}.

\section{Results of Test Cases \label{sec:results}}

Results are presented of deformational flow advection on the surface
of a sphere (\ref{subsec:advectionResults}), buoyant flow in two
and three dimensions (\ref{subsec:bouyantFlowResults}) and stably
stratified flow in two dimensions (\ref{subsec:stratifiedFlowResults}).

\subsection{Deformational Flow Advection on the Sphere \label{subsec:advectionResults}}

WWKS23 presented results of deformational flow advection on a unit
sphere \citep{LSPT12} using the adaptively implicit MPDATA on full
latitude-longitude grids, skipped latitude-longitude grids, hexagonal-icosahedra
and cubed-sphere meshes. The same grids are studied here (table \ref{tab:dx_dt})
but we use the method-of-lines advection scheme with a quasi-cubic
spatial correction. These linear advection tests all use $\gamma=1$
rather than $\gamma$ from table \ref{tab:parameterValues}. This
gives the full high-order spatial correction for all Courant numbers
and was predicted to be stable by the numerical analysis in section
\ref{subsec:NA}. Some linear advection tests are repeated with $\gamma<1$
for $c>2$, as defined as in table \ref{tab:parameterValues}, to
assess the impact on solution accuracy. The stability of using $\gamma=1$
for all Courant numbers is in contrast to the results of WWKS23 who
needed to reduce the high-order spatial correction for stability with
large Courant numbers. The other important difference from WWKS23
is that only one solver iteration is used per outer iteration (with
two outer iterations per time-step). All of the cases from WWKS23
were repeated but only select results are presented. 
\noindent \begin{center}
\begin{table}
\noindent \begin{centering}
\ %
\begin{tabular}{|l|c|c|c|c|}
\hline 
Mesh type & Nominal & N. cells & $\Delta x$ & $\Delta t$\tabularnewline
\hline 
\hline 
Latitude-longitude & $120\times60$ & 7,080 & $3.0^{o}$ & 0.01\tabularnewline
\hline 
 & $240\times120$ & 28,800 & $1.5^{o}$ & 0.005\tabularnewline
\hline 
 & $480\times240$ & 114,720 & $0.75^{o}$ & 0.0025\tabularnewline
\hline 
\hline 
Skipped latitude- & $120\times60$ & 5,310 & $3.0^{o}$ & 0.01\tabularnewline
\hline 
longitude & $240\times120$ & 21,750 & $1.5^{o}$ & 0.005\tabularnewline
\hline 
 & $480\times240$ & 88,470 & $0.75^{o}$ & 0.0025\tabularnewline
\hline 
\hline 
Cubed-sphere & C30 & 5,400 & $3.2^{o}$ & 0.01\tabularnewline
\hline 
(C$n$ is $n\times n\times6$) & C60 & 21,600 & $1.6^{o}$ & 0.005\tabularnewline
\hline 
 & C120 & 86,400 & $0.8^{o}$ & 0.0025\tabularnewline
\hline 
\hline 
Hexagonal- & HR6 & 10,242 & $2.4^{o}$ & 0.01\tabularnewline
\hline 
icosahedral & HR7 & 40,962 & $1.2^{o}$ & 0.005\tabularnewline
\hline 
 & HR8 & 163,842 & $0.6^{o}$ & 0.0025\tabularnewline
\hline 
 & HR9 & 655,362 & $0.3^{o}$ & 0.00125\tabularnewline
\hline 
\end{tabular}
\par\end{centering}
\caption{Resolutions and time-steps for deformational flow advection. $\Delta x$
is a typical cell centre to cell centre distance in degrees latitude.
The dimensionless time-step is the one that gives a nominal (global
average) Courant number of one. A complete revolution takes 5 time
units.\label{tab:dx_dt}}
\end{table}
\par\end{center}

\subsubsection{Smooth Gaussian Hill Shaped Tracer \label{subsec:smoothTracer}}

\begin{figure}
\includegraphics[width=1\textwidth]{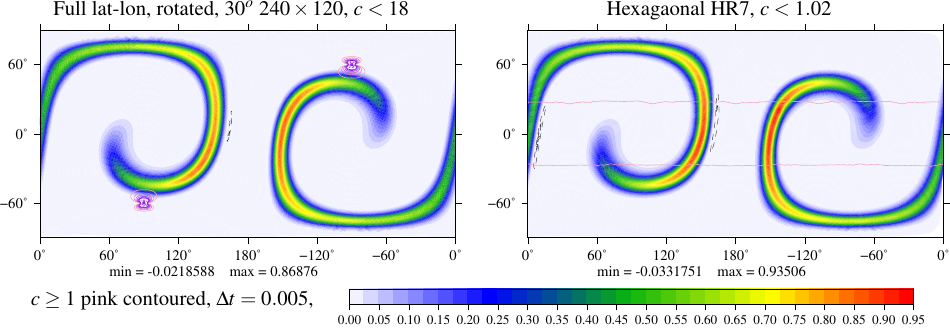}

\caption{Deformational flow on the sphere advecting Gaussian hill shaped distributions
of $\psi$ at the half-time 2.5 using $\gamma=1$. Dashed contours
at $\psi=-10^{-2},\ -2\times10^{-2},\ -3\times10^{-2}$. \label{fig:smoothDeformation}}
\end{figure}

Figure \ref{fig:smoothDeformation} shows the concentration at half
of the end time (2.5 time units) for the deformational flow advection
of two Gaussian hill shaped tracer distributions \citep{LSPT12}.
We consider the flow over the poles of a rotated latitude-longitude
grid and flow through a hexagonal icosahedral mesh using $\gamma=1$.
Pink/purple line contours show Courant numbers which are large over
the latitude-longitude poles. This test is designed to test order
of accuracy rather than boundedness so it is run without FCT and indeed
undershoots are created. The advection passes through the poles of
the grid without creating noticeable artefacts. The results are visually
identical for the latitude-longigude mesh using $\gamma$ from table
\ref{tab:parameterValues}. This is likely because the resolution
is high where the Courant number is high, so the local use of first-order
spatial discretisation does not reduce accuracy.

\noindent 

\begin{figure}
\includegraphics[width=1\textwidth]{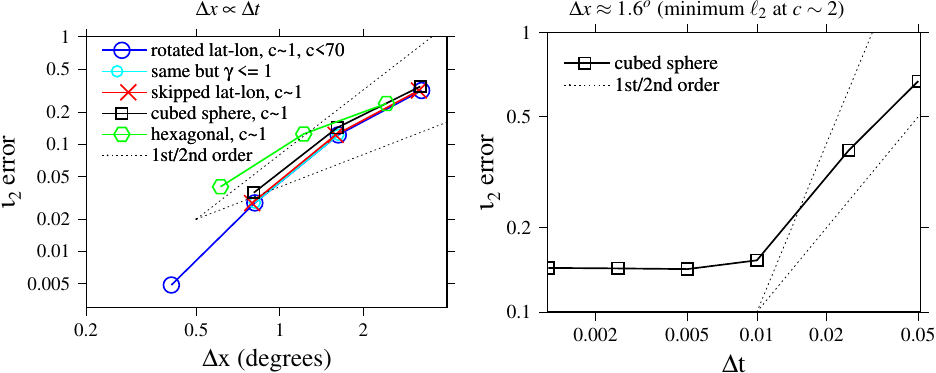}

\caption{Error norms for deformational flow advection of Gaussian hill shaped
distributions. The left panel shows errors as a function of spatial
resolution while keeping the time-step proportional. The right panel
shows errors as a function of time-step on the C60 cubed sphere. $\gamma=1$
unless otherwise stated. \label{fig:smoothDeformationErrorNorms}}
\end{figure}

The $\ell_{2}$ error norms at the end time (5 time units) are calculated
for the simulations listed in table \ref{tab:dx_dt} and are plotted
as a function of $\Delta x$ in figure \ref{fig:smoothDeformationErrorNorms}.
These simulations use $\gamma=1$ (cubic in space) apart from two
simulation on the latitude-longitude grid which use $\gamma<1$ for
$c>2$ (table \ref{tab:parameterValues}). This reduction in accuracy
where the resolution is high increases the error by about 0.2\%. The
convergence on the hexagonal grid is less than second-order. This
is because the cubic correction is not well suited to the hexagonal
grid. Second-order accurate results can be achieved on the hexagonal
grid by using a quadratic reconstruction of face values (not shown)
rather than the cubic scheme which is designed to give third-order
accuracy at regular cell centres, rather than at faces. 

When the time-step is increased while keeping spatial resolution fixed
on the cubed sphere grid (right panel of figure \ref{fig:smoothDeformationErrorNorms}),
the error remains low up to a Courant number of two, and then increases
proportional to the time-step while remaining stable, as expected
and required. 

\noindent 

\subsubsection{Slotted Cylinder \label{subsec:slottedCylinder}}

\begin{figure}
\includegraphics[width=1\textwidth]{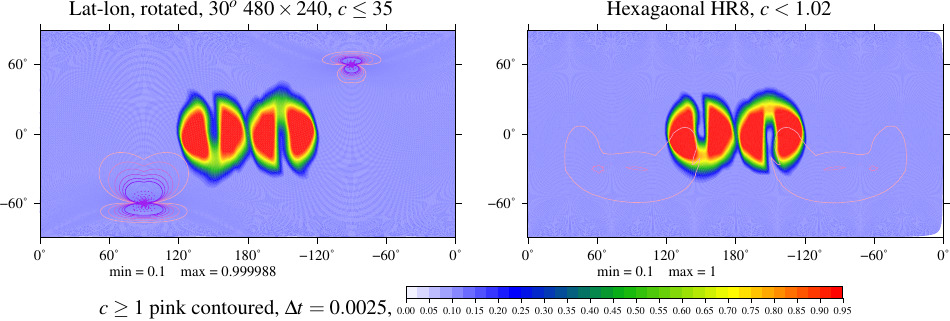}

\caption{Final results of deformational flow advection of slotted cylinders
on the sphere.\label{fig:slottedResults}}
\end{figure}

In order to test monotonicity of the first-order solution and to test
the application of FCT, a tracer field in the shape of two slotted
cylinders are advected in deformational flow \citep{LSPT12} on a
rotated latitude-longitude grid and a hexagonal icosahedra (figure
\ref{fig:slottedResults}). An initial monotonic first-order solution
is an essential part of FCT. A monotonic solution is guaranteed by
the first-order implicit in time and first-order upwind in space discretisation
(e.g. WWKS23). Figure \ref{fig:slottedResults} shows that solutions
are bounded at the final time after application of FCT, implying that
the first-order solutions are also bounded, as required. 

\noindent 

\begin{figure}
\includegraphics[width=1\textwidth]{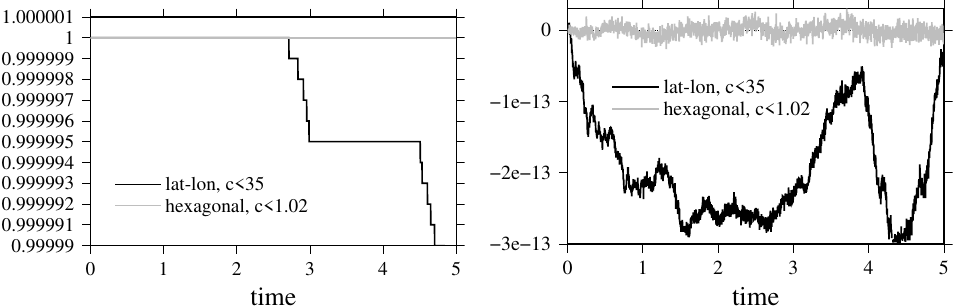}

\caption{Maximum concentration and normalised mass change during deformational
flow advection of slotted cylinders on the sphere. The minimum is
not plotted because it does not change from 0.1. The normlaised tracer
mass change is the change in mass divided by the initial tracer mass.\label{fig:slottedBoundsConservation}}
\end{figure}

The maximum tracer concentration as a function of time and the normalised
change in total tracer mass are shown in figure \ref{fig:slottedBoundsConservation}.
The minimum tracer concentration is not shown as it is always 0.1
(the initial minimum). The maximum concentration never exceeds the
initial maximum, as required. The changes in mass are of the order
$10^{-14}$ which is, importantly, machine precision rather than the
higher value of solver convergence. 

\noindent 

\subsection{Buoyancy Driven Flow \label{subsec:bouyantFlowResults}}

Positive buoyancy combined with latent heat of condensation can create
strong updrafts which, in the presence of high vertical resolution
and moderate time-steps, can lead to large vertical Courant numbers.
Tests are presented of two spatially resolved, two-dimensional dry
buoyant flows; a rising bubble \citep{BF02} and Rayleigh-B\'enard
convection \citep{SWCM22}. In order to resolve the details of convective
flows such as these, isotropic resolution is used. This gives Courant
numbers greater than one in the horizontal and vertical when combined
with moderate time-steps. These tests are used to compare fully explicit,
fully implicit and adaptive implicit time-stepping for advection,
combined with implicit treatment of acoustic and gravity waves. The
simulations use $\alpha$ and $\gamma$ as defined in table \ref{tab:parameterValues}.
The fully explicit simulations use $\beta=0$, the implicit solutions
use $\beta=1$ and the adaptive implicit (IMEX) solutions use $\beta$
from table \ref{tab:parameterValues}.

\subsubsection{Rising Bubble}

\begin{figure}
\includegraphics[width=1\textwidth]{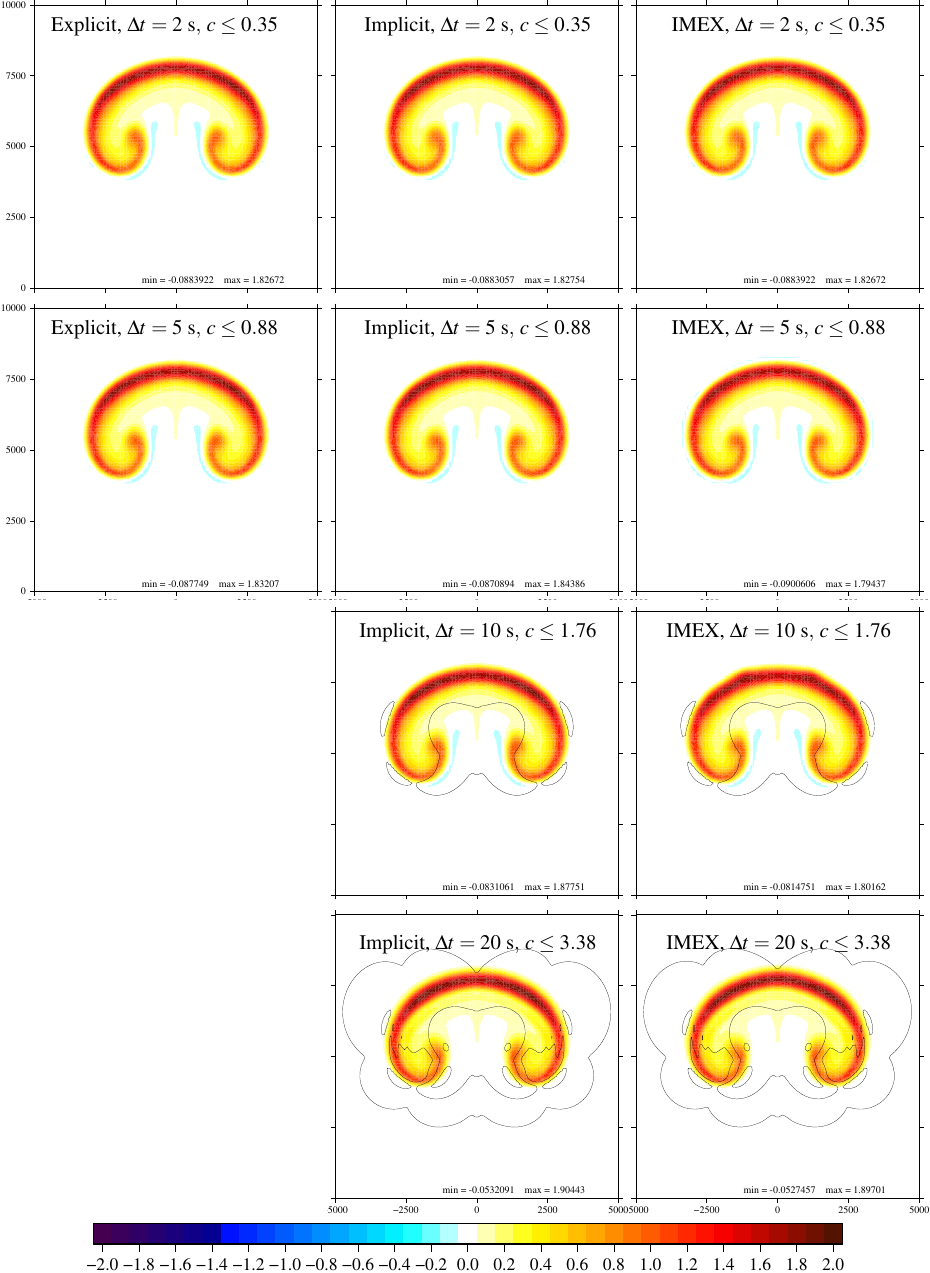}

\caption{Potential temperature anomalies (shaded, K) of the rising bubble simulation
after 1000\ s of simulation time using either explicit, implicit
or IMEX advection with different time-step sizes $\Delta t$, and
different maximum Courant numbers $c$. Courant numbers above 1.0
are line contoured in steps of 1.0. \label{fig:warmBubblesTc}}
\end{figure}

The dry rising bubble test of \citet{BF02} uses an ${x-z}$ domain
${\left[-10\thinspace\text{km},10\thinspace\text{km}\right]\times\left[0,10\thinspace\text{km}\right]}$
surrounded by zero flux boundary conditions discretised using a $\Delta x=\Delta z=100\text{ m}$
and subject gravity ${g=9.81\thinspace\text{ms}^{-2}}$. The simulation
is inviscid and initially stationary with potential temperature defined
as
\begin{eqnarray}
\theta & = & \begin{cases}
\left[300+2\cos^{2}\left(\frac{\pi r}{2}\right)\right]\thinspace\text{K} & r\le1\\
300\thinspace\text{K} & \text{otherwise},
\end{cases}\\
\text{where }r^{2} & = & \frac{\left(x-x_{c}\right)^{2}}{A_{x}^{2}}+\frac{\left(z-z_{c}\right)^{2}}{A_{z}^{2}},\\
\left(x_{c},z_{c}\right) & = & \left(0,4.5\thinspace\text{km}\right),\thinspace A_{x}=A_{z}=2\thinspace\text{km}.
\end{eqnarray}
The initial Exner pressure perturbution is zero. The potential temperature
anomalies from $300\thinspace\text{K}$ are shown in figure \ref{fig:warmBubblesTc}
after 1000\,s using time-steps from 2\,s to 20\,s. For ${\Delta t=2\thinspace\text{s}}$,
the Courant number remains below 0.35 and for ${\Delta t=5\thinspace\text{s}}$,
below 0.88. This means that the adaptively implicit (IMEX) results
are identical to the explicit results. The results using implicit
advection with small time-steps have small differences from the explicit
solutions that are difficult to see in figure \ref{fig:warmBubblesTc}.
Using ${\Delta t=10\thinspace\text{s}}$ and ${\Delta t=20\thinspace\text{s}}$
the maximum Courant numbers are 1.76 and 3.38 so the simulations using
explicit advection are unstable. At ${\Delta t=10\thinspace\text{s}}$,
the boundary between implicit and explicit time-stepping in the IMEX
simulation has initiated a disturbance in the bubble leading edge
which is unrealistic but has not lead to model instability. At ${\Delta t=20\thinspace\text{s}}$,
implicit and IMEX both have dispersion errors seen as ripples behind
the leading bubble edge, consistent with the reduced accuracy associated
with the large time-step. Therefore the adaptively implicit time-stepping
is behaving as required: enabling large stable Courant numbers locally
but without ensuring accuracy where the Courant number is large. The
test case would likely benefit from the use of FCT for advecting $\theta^{\prime}$
to avoid undershoots and ripples in $\theta^{\prime}$.

\begin{figure}
\includegraphics[width=1\textwidth]{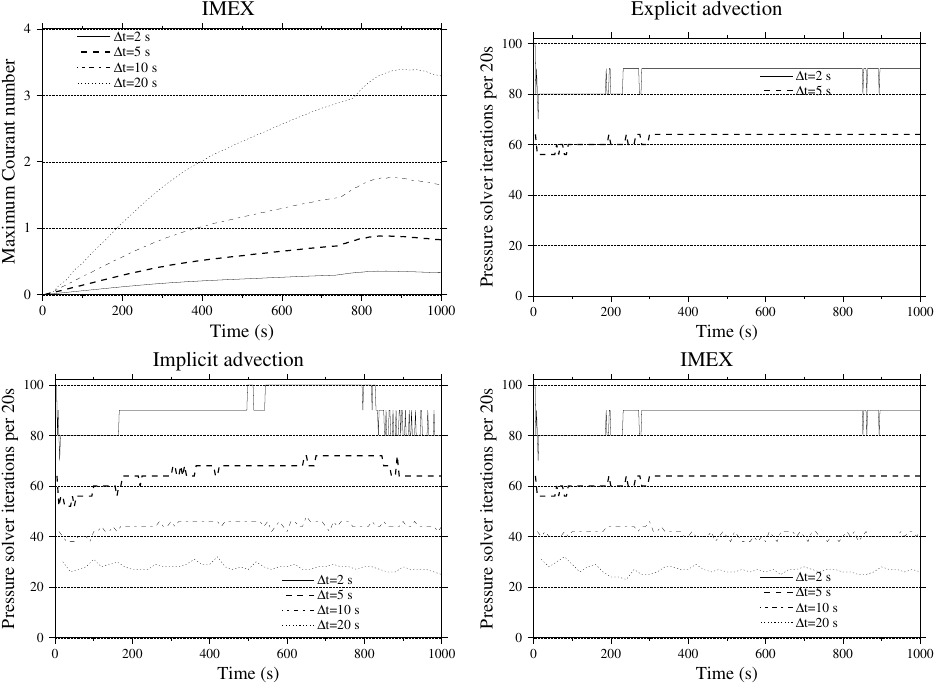}

\caption{Maximum Courant number of IMEX simulations and total number of pressure
solver iterations per 20 seconds of simulated time for the solutions
of the rising bubble using explict, implicit and IMEX advection. Increasing
the time-step always leads to fewer pressure solver iterations per
second of simulated time.\label{fig:warmBubblesCiters}}
\end{figure}

The simulations using implicit time-stepping for advection are limited
to use one solver iteration for the advection, which ensures that
the cost of the advection does not scale with the time-step. However
the larger the time-step, the more iterations are needed in the pressure
solver (to solve the Helmholtz equation for Exner pressure). Figure
\ref{fig:warmBubblesCiters} shows the maximum Courant number as a
function of time for each simulation and the number of iterations
of the pressure solver per 20\,s of simulated time (stopping criteria
defined in section \ref{subsec:HelmholtzEqn}). The bubble accelerates
throughout the simulations but the number of iterations per time-step
remains more uniform. Importantly, using a larger time-step always
results in fewer iterations in the pressure solver per 20\,s of simulated
time. For comparable time-steps, explicit and IMEX simulations use
fewer iterations than globally implicit time-stepping for advection.

\noindent 

\noindent 

\begin{figure}
\noindent \begin{centering}
\includegraphics[width=0.75\textwidth]{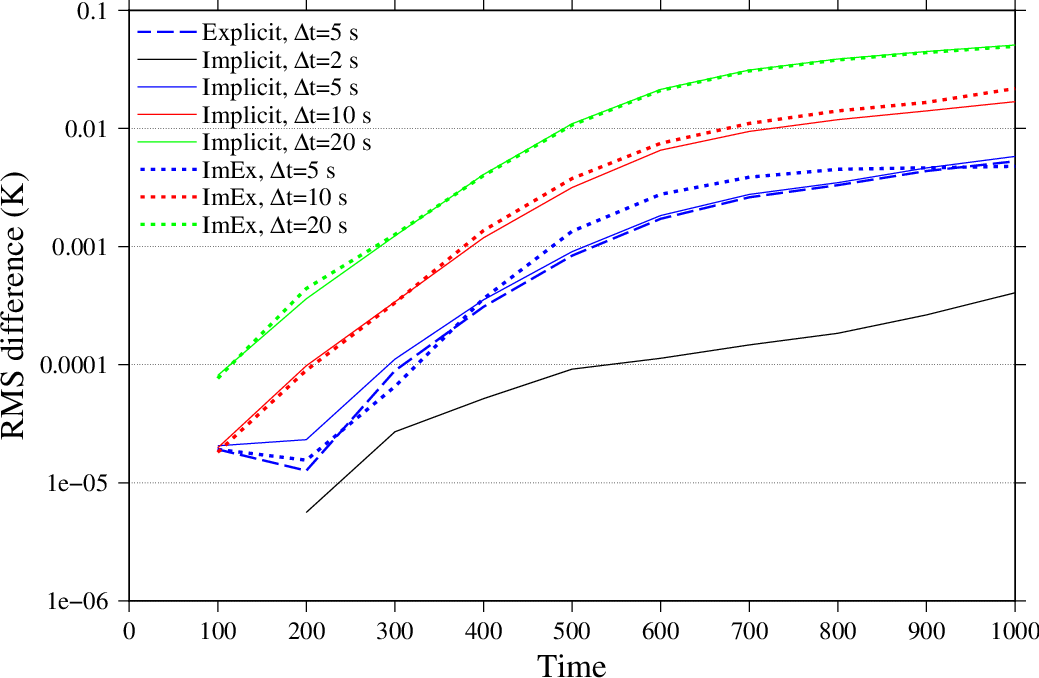}
\par\end{centering}
\caption{Root mean square differences between the temperature anomalies of
each simulation of the rising bubble and the explicit solution with
${\Delta t=2\ \text{s}}$. Implicit or IMEX advection gives similar
results to explicit for the same time-step but the larger the time-step
the greater the difference.\label{fig:warmBubblesRMSerrors}}
\end{figure}

In figure \ref{fig:warmBubblesRMSerrors}, the root mean square average
difference between the ${\Delta t=2\thinspace\text{s}}$ explicit
simulation and the other simulations are shown as a function of time.
The ${\Delta t=2\thinspace\text{s}}$ IMEX simulation results are
identical to the ${\Delta t=2\thinspace\text{s}}$ explicit results
so the zero differences is not shown on the log-scale. The ${\Delta t=2\thinspace\text{s}}$
implicit time-step results are closest to the ${\Delta t=2\thinspace\text{s}}$
explicit results. The ${\Delta t=5\thinspace\text{s}}$ results have
the next smallest RMS differences, again with the explicit and IMEX
results being identical. As the time-steps increase the RMS differences
increase and the implicit and IMEX results remain close. The adaptively
implicit advection is behaving as required -- accuracy is reduced
but large time-steps are possible.

\noindent 

\subsubsection{Rayleigh-B\'enard Convection}

Rayleigh-B\'enard convection occurs between a warm flat plate, a
distance $L_{z}$ below a cool flat plate with temperature difference
$\Delta\theta$. It is governed by two dimensionless numbers; the
Rayleigh number (the ratio of buoyancy forcing to viscous diffusion)
and the Prandtl number (the ratio of the diffusion of momentum to
the diffusion of buoyancy)
\begin{eqnarray}
\text{Ra} & = & \frac{\rho\Delta\theta\ L_{z}^{3}g}{\overline{\theta}\nu\kappa},\\
\text{Pr} & = & \frac{\nu}{\kappa}.
\end{eqnarray}
In the atmosphere $\text{Pr}=0.7$ and in the convective boundary
layer, the Rayleigh number could be $10^{16}$ \citep{SWCM22}. In
this parameter regime, convection occurs in narrow plumes, making
this a suitable test of implicit advection because Courant numbers
are only large in a small fraction of the domain. Computational constraints
mean that we simulate Rayleigh-B\'enard convection at much smaller
Rayleigh numbers in two spatial dimensions, and one simulation in
three dimensions, keeping the Rayleigh number large enough so that
convection occurs in narrow plumes.

Rayleigh-B\'enard convection is chaotic, so details of the time-stepping
will influence any snapshot of the solution, but we do not want these
details to significantly influence the solution statistics. We use
the Nusselt number (the non-dimensionalised heat transfer)
\begin{eqnarray}
\text{Nu} & = & \frac{L_{z}}{\kappa\overline{\rho}\Delta\theta}\left\langle \rho w\theta^{\prime}-\kappa\rho\frac{\partial\theta^{\prime}}{\partial z}\right\rangle _{V},
\end{eqnarray}
where $\text{\ensuremath{\left\langle \cdot\right\rangle }}_{V}$
is a volume average and $\overline{\rho}$ is the average density
in the domain.

Simulations are presented for Rayleigh numbers of $10^{6}$ and $10^{8}$
spanning the range from laminar to turbulent convection and for a
Prandtl number of one and with other parameters chosen to be in the
nearly compressible regime with round numbers (table \ref{tab:RBparams}).
SI units are given in table \ref{tab:RBparams} but are not used for
Rayleigh-B\'enard convection throughout the rest of this section
as they have no physical significance. The resolution, domain size
and run length requirements are taken from \citet{SWCM22}. The spatial
and temporal resolutions are in tables \ref{tab:RBparams} and \ref{tab:RBresolution}.

\noindent 
\begin{table}
\begin{tabular}{|l|c|c|c|c|}
\hline 
Parameter &  &  &  & units\tabularnewline
\hline 
\hline 
Rayleigh number & Ra & $10^{6}$ & $10^{8}$ & --\tabularnewline
\hline 
Gravitational acceleration & $g$ & 10 & 10 & m s\textsuperscript{-2}\tabularnewline
\hline 
Domain height & $L_{z}$ & 1 & 1 & m\tabularnewline
\hline 
Domain width & $L_{x}$, $L_{y}$ & 2.02 & 2.02 & m\tabularnewline
\hline 
Top to bottom temperature difference & $\Delta T$ & 1 & 1 & K\tabularnewline
\hline 
Average temperature & $\overline{T}$ & 1000 & 1000 & K\tabularnewline
\hline 
Kinematic viscosity & $\nu$ & $10^{-4}$ & $10^{-5}$ & m s\textsuperscript{-2}\tabularnewline
\hline 
Gas constant & R & 100 & 100 & J kg\textsuperscript{-1} K\textsuperscript{-1}\tabularnewline
\hline 
Total simulation time &  & 1000 & 1000 & s\tabularnewline
\hline 
Time-step & $\Delta t$ &  & 0.2 & s\tabularnewline
\hline 
Grid spacing & $\Delta x$, $\Delta z$ &  & 0.01 & m\tabularnewline
\hline 
\end{tabular}

\caption{Parameters for simulating Rayleigh-B\'enard convection with Prandtl
number one. Grid spacing, time-steps and maximum Courant numbers for
the ${\text{Ra}=10^{6}}$ simulations are in table \ref{tab:RBresolution}.\label{tab:RBparams}}
\end{table}

\noindent 
\begin{table}
\noindent \begin{centering}
\begin{tabular}{|c|c|c|c|c|}
\hline 
\backslashbox{$\Delta t$}{$\Delta x,\ \Delta z$} & 0.01 & 0.02 & 0.02 (3D) & 0.04\tabularnewline
\hline 
\hline 
0.05 & 0.51 &  &  & \tabularnewline
\hline 
0.1 &  & 0.40 &  & 0.19\tabularnewline
\hline 
0.2 &  & 0.79 &  & 0.39\tabularnewline
\hline 
0.5 &  & 2.0 &  & 0.97\tabularnewline
\hline 
1 &  & 4.1 & 5.4 & 1.9\tabularnewline
\hline 
2.5 &  & 11.3 &  & 4.8\tabularnewline
\hline 
5 &  &  &  & 8.6\tabularnewline
\hline 
\end{tabular}
\par\end{centering}
\caption{Maximum Courant numbers for simulations of Rayleigh-B\'enard convection
with ${\text{Ra}=10^{6}}$ at different resolutions in two- and three-dimensions.\label{tab:RBresolution}}
\end{table}

\begin{figure}
\noindent \begin{centering}
\includegraphics[width=0.95\textwidth]{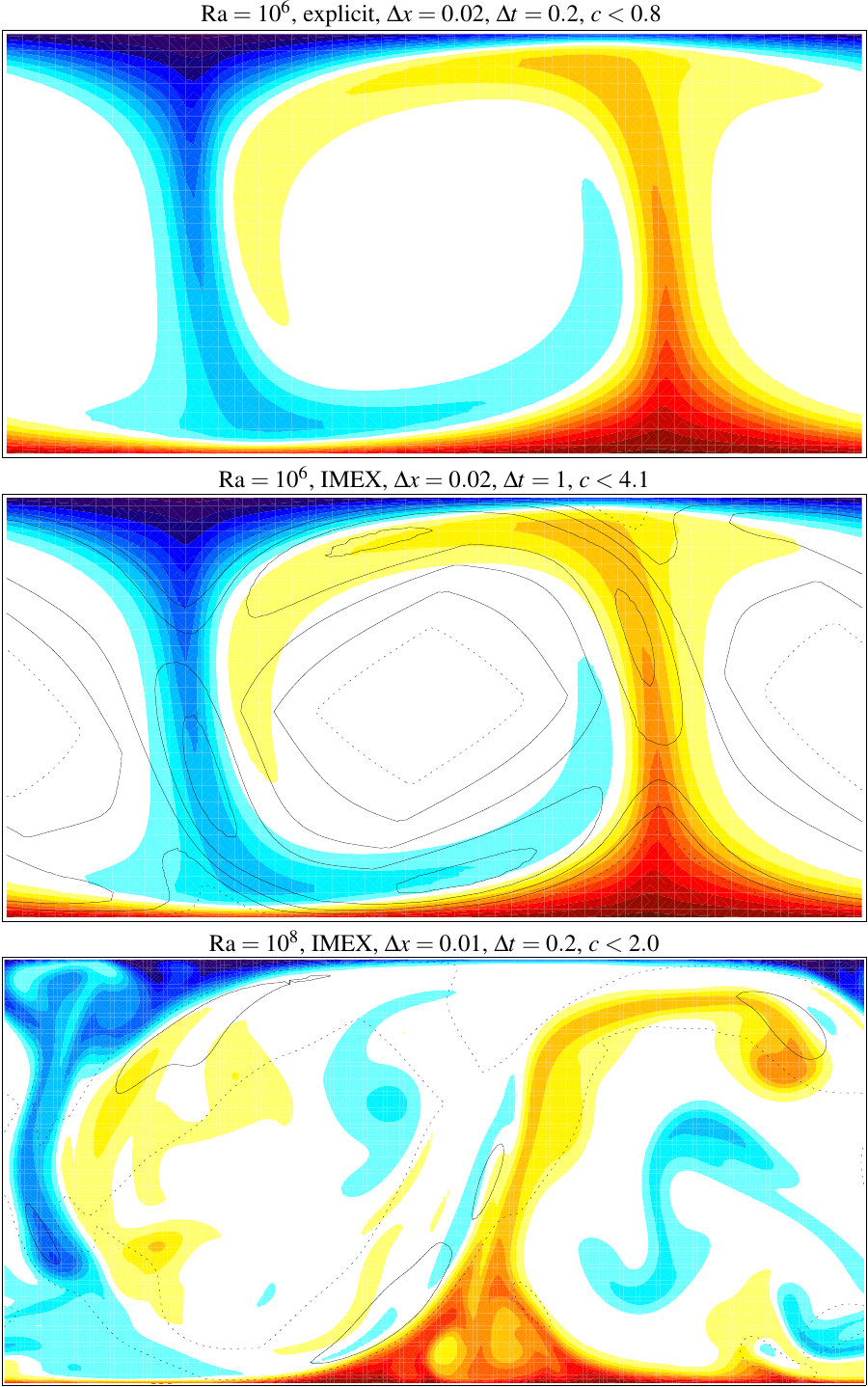}
\par\end{centering}
\caption{Potential temperature perturbations of Rayleigh-B\'enard convection
after 1000 time units calculated using explicit and IMEX advection
with different time-steps. The dotted contour show line indicates
the Courant number ${c=1}$, and larger Courant numbers are shown
every 0.5. \label{fig:RBTc}}
\end{figure}

Potential temperature perturbations at $t=1000$ time units are shown
in figure \ref{fig:RBTc} for two $\text{Ra}=10^{6}$ simulations
and the $\text{Ra}=10^{8}$ simulation. The $\text{Ra}=10^{6}$ simulations
are laminar but not steady and the $\text{Ra}=10^{8}$ simulation
is chaotic with some turbulent eddies. The snapshot at $\text{Ra}=10^{8}$
is similar to that of \citet[their fig 1b]{SWCM22} but their simulation
is at higher spatial and temporal resolution. Contours showing where
the Courant number is greater than one are overlayed in figure \ref{fig:RBTc}.
There are no obvious artefacts where time-stepping switches from explicit
to implicit and, for $\text{Ra}=10^{6}$, the small time-step, explicit
solution is similar to the larger time-step IMEX solution, despite
the lower temporal accuracy where the Courant number reaches 4.1.
A repeated IMEX simulation using $\Delta x=0.02$, $\Delta t=1$ but
with $\gamma=1$ for all $c$ gives solutions that look identical
to the middle plot of \ref{fig:RBTc}, with a root mean square difference
of 12\%, despite the potential for chaotic growth of differences. 

\noindent 

\begin{figure}
\noindent \begin{centering}
\includegraphics[width=0.95\textwidth]{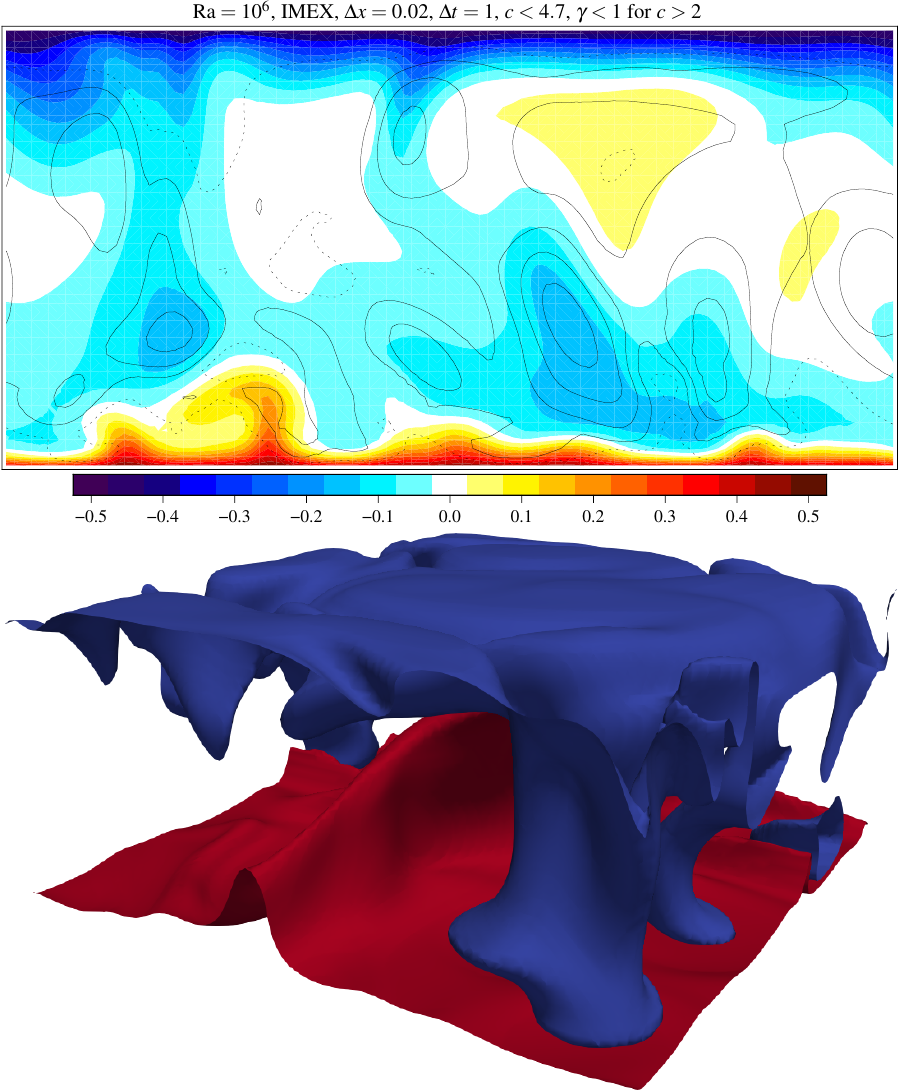}
\par\end{centering}
\caption{Potential temperature perturbations at 1000 time units from the 3D
simulation Rayleigh-B\'enard convection. The top panel shows a cross
sections with the dotted contour line indicating the Courant number
${c=1}$, and larger Courant numbers are contoured every 0.5. The
bottom panel shows iso-surfaces of $\theta^{\prime}=\pm0.1$. \label{fig:RBT_3d}}
\end{figure}

Results of a 3D simulation of Rayleigh-B\'enard convection are shown
in figure \ref{fig:RBT_3d}. This uses the parameters for $\text{Ra}=10^{6}$
with $\Delta x=\Delta y=0.02$ and $\Delta t=1$ giving a maximum
Courant number of 4.7. The simulation in figure \ref{fig:RBT_3d}
uses $\gamma<1$ for $c>2$. Isosurfaces of $\theta^{\prime}=\pm0.1$
are shown in the bottom of figure \ref{fig:RBT_3d}. This demonstrates
that exactly the same algorithms work in 3D, although the plumes are
less vertically coherent. 

\noindent 

\begin{figure}
\noindent \begin{centering}
\includegraphics[width=0.9\textwidth]{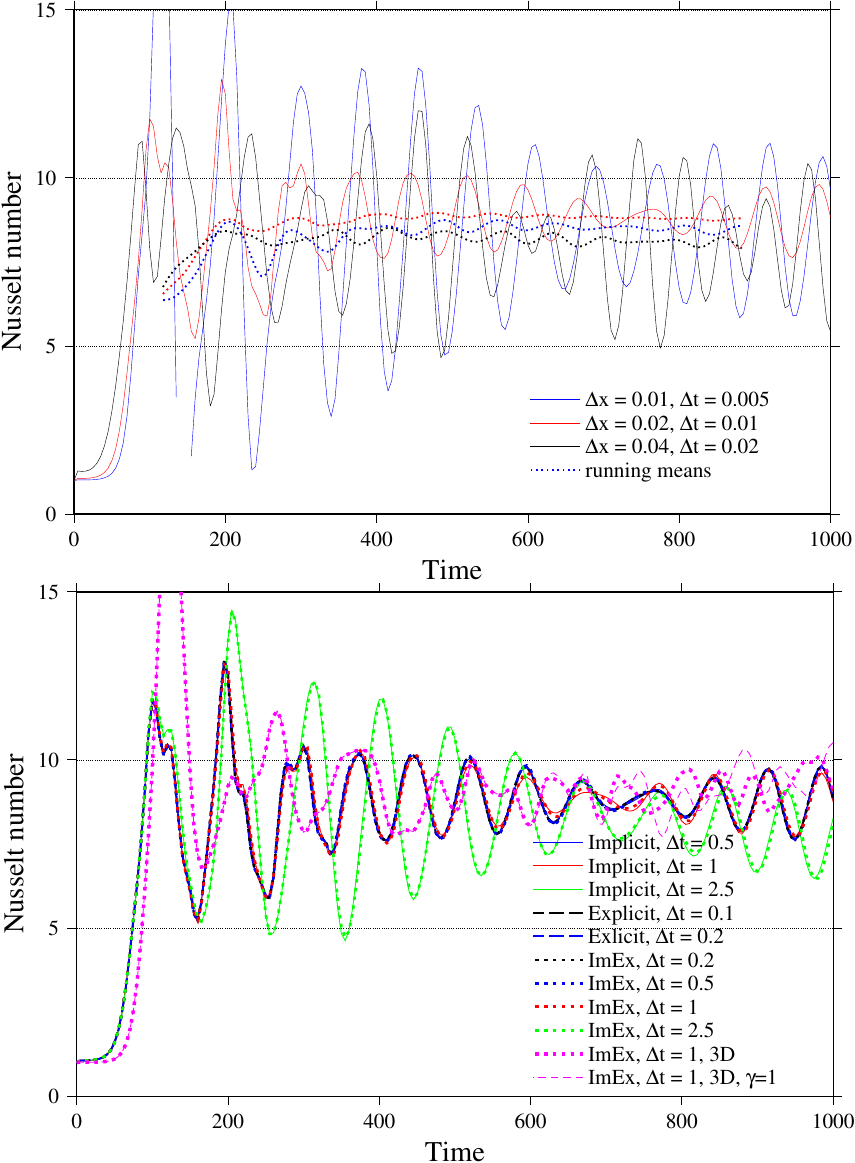}
\par\end{centering}
\caption{The top panel shows the domain averaged Nusselt number for simulations
of 2D Rayleigh-B\'enard convection with ${\text{Ra}=10^{6}}$ using
explicit advection and different spatial resolutions and time-step
scaling with spatial resolution. The continuous lines are domain averages
every 5 time units. The dotted lines are averaged over 20 sets of
5 time units. The bottom panel shows Nusselt numbers averaged over
the top and bottom boundaries of 2D and 3D domains every 5 time units
for simulations all using spatial resolution of 0.02. \label{fig:Nusslets}}
\end{figure}

Rayleigh-B\'enard convection is chaotic so in order to assess the
impact of adaptive implicit time-stepping and large time-steps, we
compare the evolution of the Nusselt number between different simulations.
The top panel of figure \ref{fig:Nusslets} shows the domain average
Nusselt number for $\text{Ra}=10^{6}$ for three simulations of two
dimensional convection using different spatial resolutions, and using
explicit advection. The time-step scales with spatial resolution so
that the maximum Courant number is around 0.5. All three resolutions
show a spin up followed by oscillations in Nusselt number as the plume
meanders. The dotted lines are time averages. The highest resolution
produces the middle value of Nusselt number and the largest oscillations.
The simulations are not converged but have similar behaviour. 

The middle spatial resolution (${\Delta x=0.02}$) is used to compare
time-step size and the impact of adaptive implicit advection. The
bottom panel of figure \ref{fig:Nusslets} shows the Nusselt number
averaged over the top and bottom boundaries for ${\Delta x=0.02}$.
The Nusselt number averaged just at the boundaries oscillates less
than the fluid averaged Nusselt number so time averages are not needed.
The spin up and time average Nusselt number is not sensitive to time-step
or time-stepping method while the amplitude and period of the oscillations
is senstive to the time-step but much less sensitive to the time-stepping
method. For time-steps of ${\Delta t=2.5}$ time units and larger,
(${c>10}$), the phase and period of the oscillations is noticeably
different from the smaller time-step simulations. The Nusselt numbers
of the 3D simulations are also displayed, in the lower panel of figure
\ref{fig:Nusslets}, showing less regular oscillations but a similar
mean Nusselt number and similar magnitude oscillations. 3D simulations
with $\gamma\le1$ and with $\gamma=1$ compared in figure \ref{fig:Nusslets}
confirm the small impact of using $\gamma<1$ locally. 

Again, the adaptive implicit time-stepping is behaving as required,
enabling Courant numbers significantly larger than one in a fraction
of the domain, without detrimental effects on accuracy.

\noindent 

\begin{figure}
\includegraphics[width=1\textwidth]{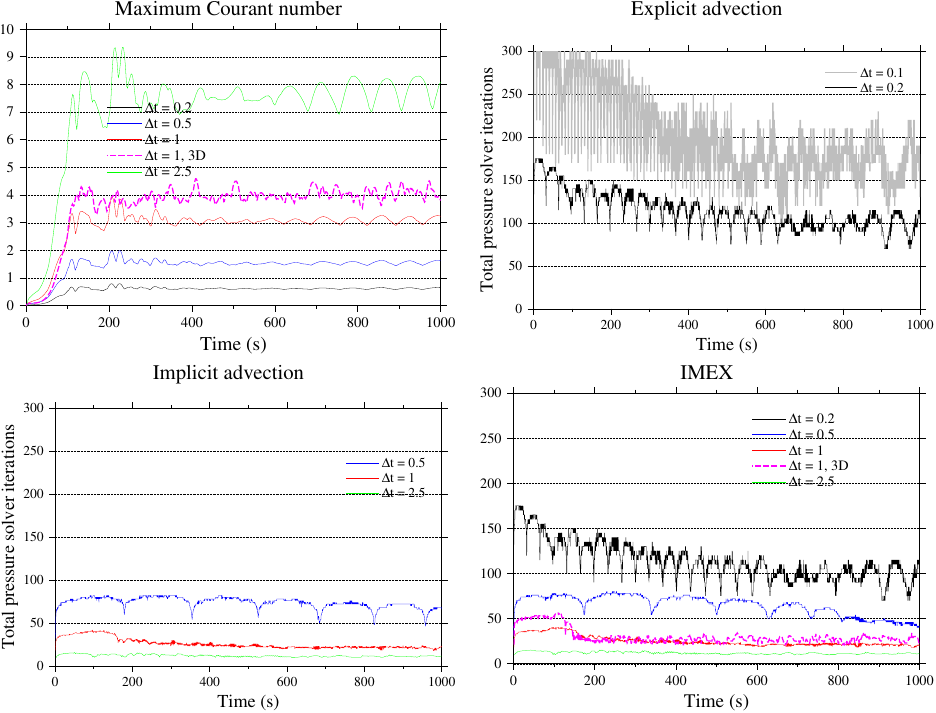}

\caption{Maximum Courant number and total number of pressure solver iterations
per unit simulated time for the solutions of Rayleigh-B\'enard convection
with ${\text{Ra}=10^{6}}$ using ${\Delta x=0.02}$.\label{fig:RBCiters}}
\end{figure}

The maximum Courant number as a function of time for the adaptively
implicit (IMEX) solutions for a range of time-steps is shown in figure
\ref{fig:RBCiters}, along with the number of iterations of the pressure
solver per unit time for each time-stepping type and each time-step,
all for the middle spatial resolution ($\Delta x=0.02$). The 3D simulation
has larger maximum Courant numbers than the 2D simulation with the
same time-step, due to the stronger updrafts in the plumes.

For each time-stepping type, the number of pressure solver iterations
per unit time reduces as the time-step increases, indicating that
implicit and adaptively implicit time-stepping is cost effective for
increasing time-steps. The 3D IMEX simulation uses slightly more iterations
that the corresponding 2D simulation, likely because the maximum Courant
numbers are higher.

\noindent 

\subsection{Stably Stratified Flow \label{subsec:stratifiedFlowResults}}

Strongly stratified flow introduces an additional complexity because
gravity waves can be fast, so they are also treated implicitly. Two
two-dimensional test cases with stratification are used; strongly
stratified, near hydrostatic flow over a 1\,m hill \citep{MDW+10}
and flow over a mountain range \citep{SLF+02}. 

\subsubsection{Flow over a 1\,m hill}

This test case is defined by \citet{MDW+10} and is usually run with
$N\Delta t<1$ where $N$ is the Brunt--V\"{a}is\"{a}l\"{a} frequency.
$N\Delta t=1$ is the time-step limit for treating gravity waves explicitly
using forward-backward or RK2 time-stepping. This test has strong
stratification so that the time-step can be increased to give $N\Delta t>1$
in order to demonstrate the benefits of implicit treatment of gravity
waves.

The hill has profile 
\begin{equation}
h\left(x\right)=\frac{a^{2}}{x^{2}+a^{2}},
\end{equation}
where $a=10\thinspace\text{km}$ in a domain 50\,km deep and 240\,km
wide (centred on $x=0$) with rigid top and bottom boundaries and
periodic left and right boundaries. A wave absorbing layer is applied
in the top 20\,km with damping coefficient $\overline{\mu}\Delta t=0.3$
(term $-\overline{\mu}w$ is added to the RHS of the $w$ equation).
The grid spacing is $\Delta x=2\thinspace\text{km}$ and $\Delta z=250\thinspace\text{m}$.
The initial wind is $20\thinspace\text{m/s}$ in the $x$-direction
and the initial temperature is uniformly $250\thinspace\text{K}$
with the Exner pressure in discrete hydrostatic balance. The time-steps
and resulting maximum Courant numbers and $N\Delta t$ for simulations
which all use adaptively implicit (IMEX) time-stepping for advection
are shown in table \ref{tab:hillTimeSteps}.

\noindent 
\begin{table}
\noindent \begin{centering}
\begin{tabular}{|c|c|c|}
\hline 
$\Delta t$ (s) & max $c$ & $N\Delta t$\tabularnewline
\hline 
\hline 
20 & 0.2 & 0.39\tabularnewline
\hline 
40 & 0.4 & 0.78\tabularnewline
\hline 
100 & 1 & 1.95\tabularnewline
\hline 
200 & 2 & 3.92\tabularnewline
\hline 
500 & 5 & 9.80\tabularnewline
\hline 
1000 & 10 & 19.6\tabularnewline
\hline 
\end{tabular}
\par\end{centering}
\caption{Time-steps and resulting maximum Courant numbers and $N\Delta t$
for simulations of flow over a 1\,m hill using adaptively implicit
(IMEX) time-stepping for advection.\label{tab:hillTimeSteps}}
\end{table}

\begin{figure}
\noindent \begin{centering}
\includegraphics[width=0.8\textwidth]{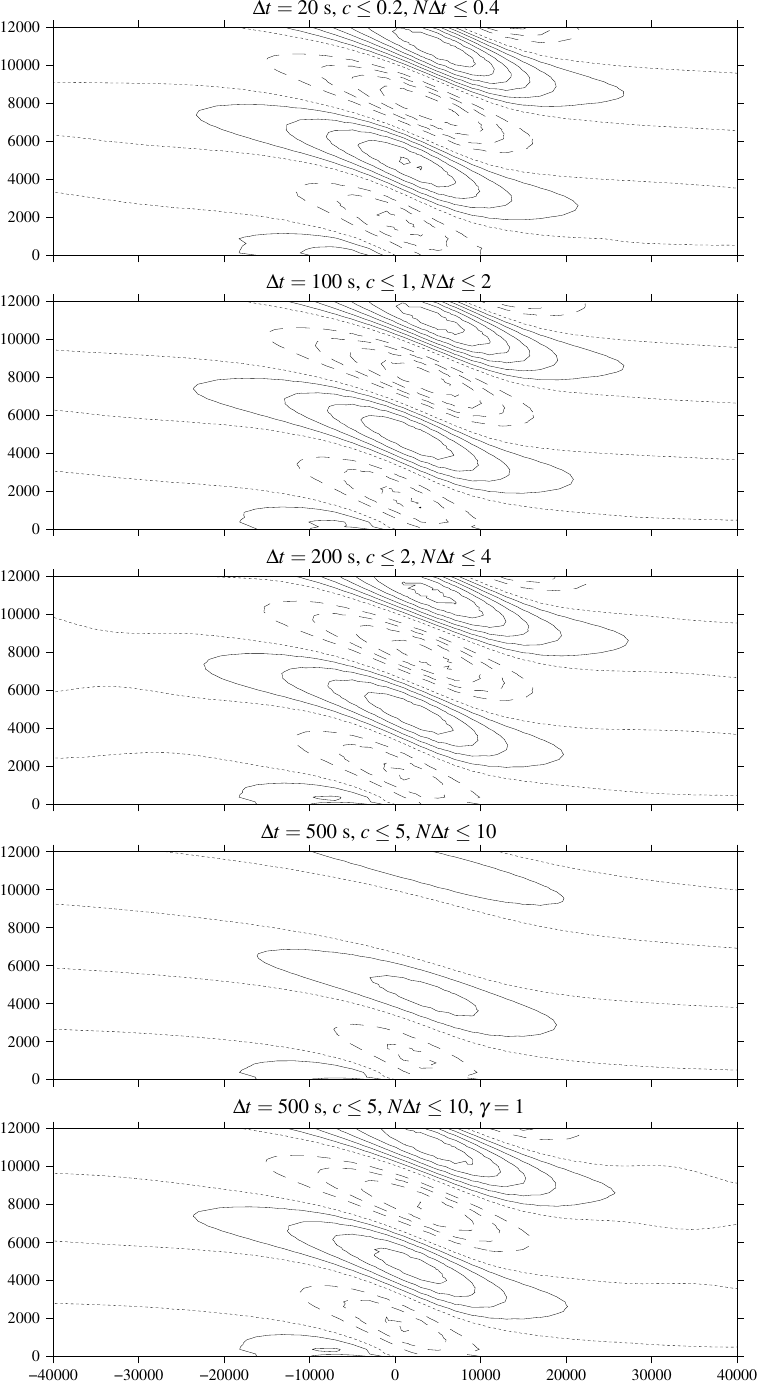}
\par\end{centering}
\caption{Vertical velocity after $1.5\times10^{4}\text{ seconds}$ for stably
stratified flow over a 1m hill. Contours every ${5\times10^{-4}\text{ms}^{-1}}$,
zero contour dotted. As the time-step increases, the accuracy reduces
but stability is maintained since $\gamma<1$ for $c>2$.\label{fig:1mhillw}}
\end{figure}

The vertical velocity after $15,000\thinspace\text{s}$ for time-steps
between $\Delta t=20\thinspace\text{s}$ and $\Delta t=500\thinspace\text{s}$
are shown in figure \ref{fig:1mhillw} using $\gamma<1$ for $c>2$
as defined in table \ref{tab:parameterValues}. For Courant numbers
up to two, the results are similar to those of, for example, \citet{MDW+10}.
As the time-step increases beyond $c=2$, the waves decrease in amplitude
due to the reduced high-order spatial correction ($\gamma<1$). The
adaptively implicit time-stepping is behaving as required, stabilising
the solution for large time-steps. This simulation is stable and accurate
for $\Delta t=500\ \text{s}$ when using $\gamma=1$ (bottom panel
of figure \ref{fig:1mhillw}), but this is not recommended due to
spurious results in the Sch\"ar mountain range case (section \ref{subsec:Schar}).
The simulation is unstable for $\Delta t=1000\thinspace\text{s}$
which is due to a combination of the large horizontal Courant number
and large stratification. Simulations with time-steps this large can
be stabilised by using more outer iterations per time-step but this
increases cost and so these results are not presented.

\noindent 

\begin{figure}
\noindent \begin{centering}
\includegraphics[width=0.75\textwidth]{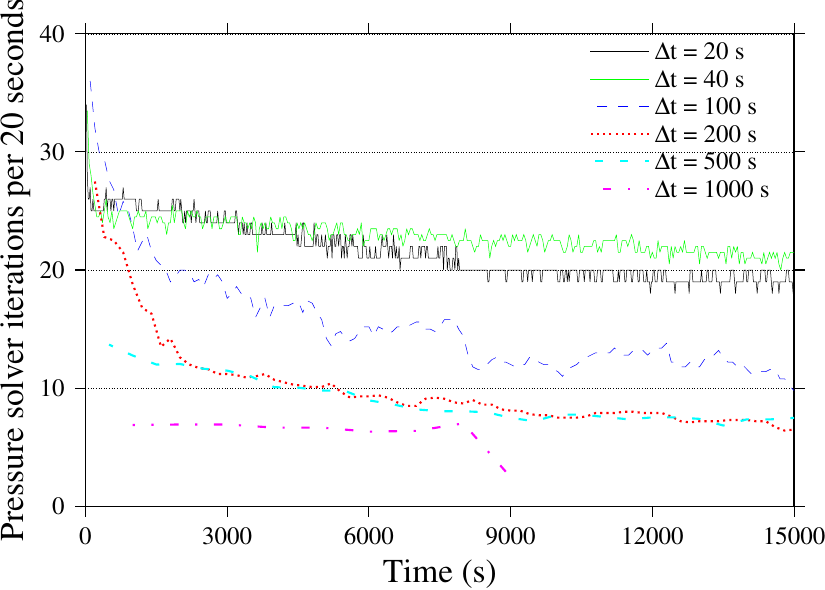}
\par\end{centering}
\caption{Total number of pressure solver iterations per 20 seconds of simulated
time for the IMEX simulations of stratified flow over a 1m hill. Taking
larger time-steps means fewer iterations in the pressure solver in
total. The simulation with the largest time step ${\Delta t=1000\ \text{s}}$
is not stable when using just two outer iterations per time step.\label{fig:1mhilliterations}}
\end{figure}

The number of pressure solver iterations per 20 seconds of simulated
time for simulations using each time-step is shown in figure \ref{fig:1mhilliterations}.
As with the simulations of buoyant flows, increasing the time-step
decreases the number of pressure solver iterations per 20 seconds
of simulated time (although the number of pressure iterations per
time-step increases) demonstrating the cost-effectiveness of implicit
advection.

\noindent 

\subsubsection{Flow over Sch\"ar mountain range\label{subsec:Schar}}

The \citet{SLF+02} test with flow over a mountain range is useful
to test the adaptive implicit time-stepping through grid cells that
are distorted due to orography, which effectively increases the Courant
number as defined in (\ref{eq:c}). The mountain range is defined
as
\begin{eqnarray}
h\left(x\right) & = & h_{m}\exp\left[-\left(\frac{x}{a}\right)^{2}\right]\cos^{2}\frac{\pi x}{\lambda},\\
h_{m} & = & 250\ \text{m},\ a=5\ \text{km},\ \lambda=4\ \text{km}.
\end{eqnarray}
Following \citet{MDW+10} we use a domain of $330\text{ km}\times100\text{ km}$,
$\Delta x=\Delta z=0.5\text{ km}$, and an absorbing layer in the
top 10 km of the domain with damping coefficient $\overline{\mu}\Delta t=1.2$.
A basic terrain following grid is used, as defined in \citet{Kle11},
with periodic boundary conditions in the $x$-direction and zero flux
boundary conditions at the top and bottom. The air is frictionless
and the initial conditions consist of a wind of $10\text{ ms}^{-1}$
in the $x$-direction, a surface temperature of $288\thinspace\text{K}$
and stratification with $N=0.01\text{ s}^{-1}$ in hydrostatic balance.

\begin{figure}
\noindent \begin{centering}
\includegraphics[width=0.8\textwidth]{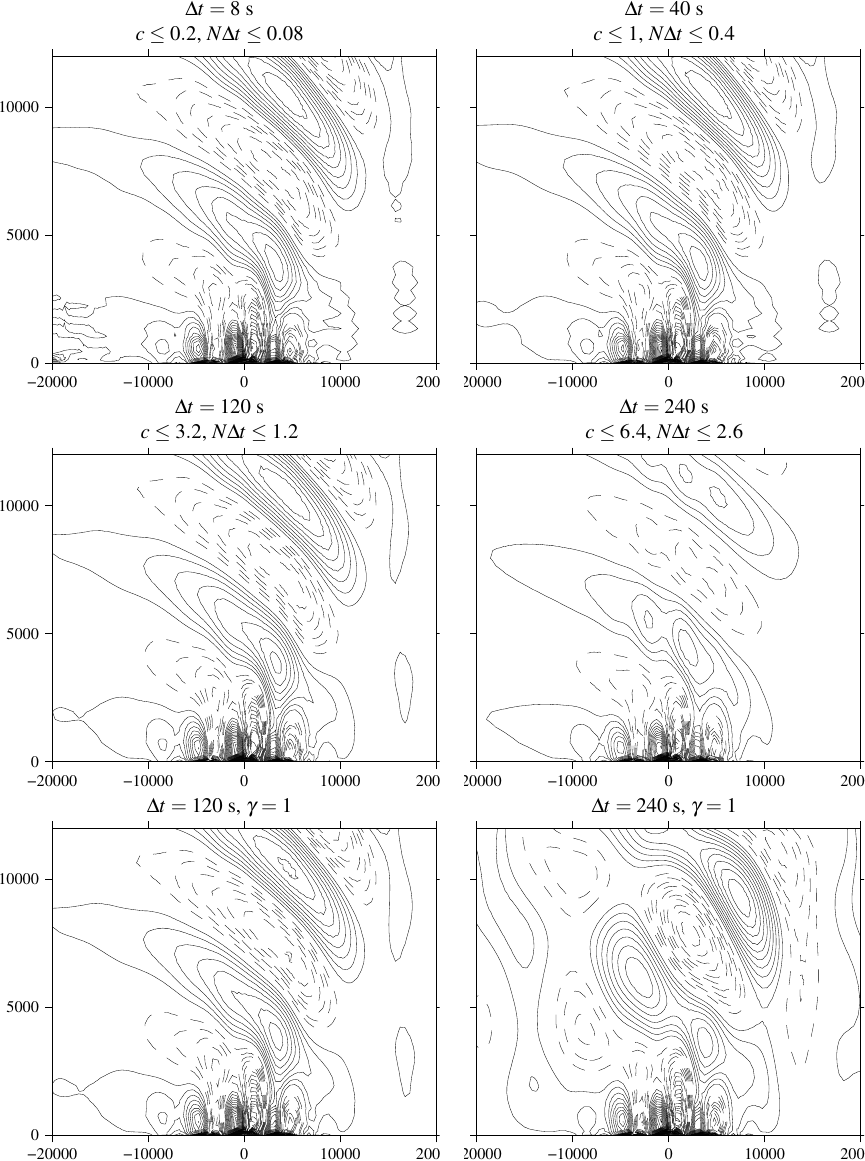}
\par\end{centering}
\caption{Vertical velocity after ${1.8\times10^{4}\text{ seconds}}$ for flow
over Sch\"{a}r mountains using IMEX advection. Contours every $0.05\ \text{ms}^{-1}$.
Grid-scale oscillations associated with this particular co-located
method contaminate the ${\Delta t=40\ \text{s}}$ and ${\Delta t=8\ \text{s}}$
results. \label{fig:scharw}}
\end{figure}

\citet{MDW+10} use a time-step of 8\ s for this test, giving a Courant
number of 0.2. Our simulations use time-steps of 8\,s, 40\,s, 120\ s
and 240\ s with adaptively implicit time-stepping. Contours of vertical
velocity are shown in figure \ref{fig:scharw}. The 120\ s and 240\ s
solutions have $N\Delta t>1$ and so are only stable because gravity
waves are treated implicitly. The standard simulations use $\gamma<1$
for $c>2$ as defined in table \ref{tab:parameterValues}. For the
two smaller time-steps, grid scale oscillations are evident downstream
of the mountain, related to the co-location of pressure and velocity
in this spatial discretisation. These oscillations are damped when
using the two larger time-steps. This is consistent with the pressure
velocity algorithm which uses a compact molecule for the pressure
equation \citep{RC83} which damps the computational mode effectively
when fast waves are not resolved. As the time-step increases beyond
$c=1$ and beyond $N\Delta t=1$, the mountain waves become damped.
This test can be run stably with $\gamma=1$ (full cubic spatial correction
for all Courant numbers), giving the results in the bottom panels
of figure \ref{fig:scharw}. At $\Delta t=240\ \text{s}$, spurious
features are visible where the undamped gravity waves generated by
the mountain interact with the acoustic waves that circulate the domain,
forced by the initial condition shock. Spurious features such as these
are the reason for using $\gamma<1$ for $c>2$.

\noindent 

\begin{figure}
\noindent \begin{centering}
\includegraphics[width=0.75\textwidth]{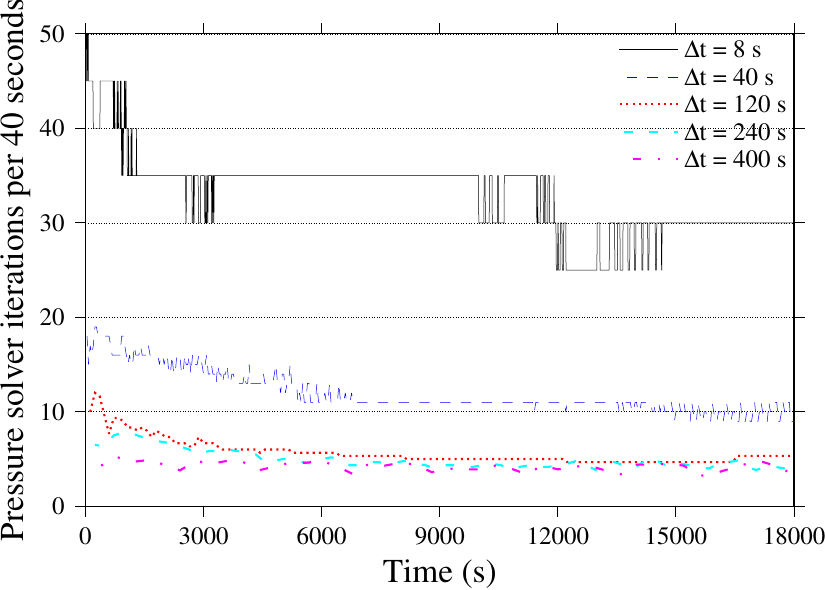}
\par\end{centering}
\caption{Total number of pressure solver iterations per 40 seconds of simulated
time for the IMEX simulations of flow over the Sch\"ar mountain range.
Taking larger time-steps means fewer iterations in the pressure solver
in total.\label{fig:scharIterations}}
\end{figure}

The number of pressure solver iterations per 40 seconds of simulated
time for the adaptively implicit solution of flow over the Sch\"ar
mountain range are shown in figure \ref{fig:scharIterations}. Taking
larger time-steps means fewer iterations in the pressure solver in
total, as with the other test cases. 

\noindent 

\section{Summary and Conclusions \label{sec:concs}}

Adaptively implicit advection uses implicit time-stepping where the
Courant number is above one (or thereabouts) and standard explicit
time-stepping elsewhere. We presented an adaptively implicit advection
scheme with the following properties:
\begin{enumerate}
\item Accuracy and cost comparable with explicit time-stepping where the
Courant number is less than 1 (in practice, we use a threshold of
0.8).
\item Second- or third-order accuracy for Courant numbers up to 2.
\item Stable and at least first-order accurate for Courant numbers larger
than 2. The advection scheme is provably monotonic for all Courant
numbers and this is demonstrated with linear advection test cases. 
\item Stable solutions with large Courant numbers are also achieved solving
the fully compressible Navier-Stokes equations using a combination
of implicit treatment of gravity and acoustic waves with the adaptively
implicit advection.
\item Given the availability of a first-order monotonic solution, FCT can
be used with the adaptively implicit advection. In the methods-of-lines
approach, a quasi-cubic upwind spatial discretisation is combined
with the two-stage implicit-explicit Runge-Kutta time stepping using
off-centring and implicitness that varies in space and time.
\item The adaptively implicit advection is suitable for a variety of mesh
structures. It is demonstrated on cubed-sphere, hexagonal and latitude-longitude
meshes of the sphere and terrain-following meshes over orography.
\item Exact local conservation is achieved by defining parameters such as
the spatially-varying temporal off-centring at faces between grid
cells, and ensuring that these spatially varying parameters are inside
the divergence of the transport term. 
\item Mutli-tracer efficiency has not been demonstrated, but should be possible
with adpatively implicit advection because the same preconditioner
can be used for each tracer being transported, after which, only one
solver iteration is needed for transporting each tracer.
\item Good parallel scaling has not been demonstrated here, but again, should
be possible because the implicit parts of the solution use only nearest
neighbour communication, since order of accuracy is reduced to first-order
where the Courant number is large. 
\item Adaptively implicit advection is demonstrated to be effective using
just one solver iteration per application of the advection operator.
Therefore, the cost of the advection does not scale with the Courant
number. When used with a solution of the Euler or Navier-Stokes equations,
more iterations are needed every time-step to solve the Helmholtz
(pressure) equation for large time-steps. However the number of iterations
to solve the pressure equation scales slower than linearly with the
Courant number so cost savings are always made when the time-step
is increased.
\end{enumerate}
In conclusion, adaptively implicit time-stepping for advection, including
in the fully compressible Navier-Stokes equations, exhibits efficiency,
stability and accuracy where needed across a range of test cases.
Further research is needed to explore parallel scaling aspects and
application in the context of other 3D tests, as well as realistic
weather and climate models.

\section*{Open Research and Code Availability}

The code used to produce these results is at \url{https://zenodo.org/doi/10.5281/zenodo.593780}
\citep{AtmosFOAM24}. The test cases set ups are at \url{https://zenodo.org/doi/10.5281/zenodo.11546593}
\citep{AtmosFOAM_run24}. The code can be compiled with OpenFOAM 11
\citep{OpenFOAM24} \url{https://openfoam.org/version/11/}. 

\bibliography{numerics}

\appendix

\section{Mathematical Notation\label{appx:notation}}

The notation used in the paper is listed in table \ref{tab:notation}.
Some notation is only used in individual sections and defined close
to where it is used and so is not repeated in table \ref{tab:notation}.

\noindent 
\begin{table}
\begin{tabular}{|l|c|c|l|c|}
\hline 
\textbf{Physical Properties} &  &  &  & \tabularnewline
\hline 
Tracer  & $\psi$ &  & Source term of $\psi$ & $S_{\psi}$\tabularnewline
\hline 
Density & $\rho$ &  & Diffusion coefficient & $K$\tabularnewline
\hline 
Velocity & $\mathbf{u}$ &  & Thermal diffusivity & $\kappa$\tabularnewline
\hline 
Vertical velocity component & $w$ &  & Kinematic viscosity & $\nu$\tabularnewline
\hline 
Horizontal velocity & $\mathbf{v}$ &  & Exner pressure & $\pi$\tabularnewline
\hline 
Potential temperature & $\theta$ &  & Gas constant & $R$\tabularnewline
\hline 
Nusselt number & Nu &  & Brunt--V\"{a}is\"{a}l\"{a} frequency & $N$\tabularnewline
\hline 
Rayleigh number & Ra &  & Acceleration due to gravity & $\mathbf{g}$\tabularnewline
\hline 
\textbf{Discretisation related} &  &  & Courant number & $c$\tabularnewline
\hline 
Mass flux & $\phi$ &  & Amplification factor & $\mathcal{A}$\tabularnewline
\hline 
Volume flux & $U_{f}$ &  & Normalised RMS error norm & $\ell_{2}$\tabularnewline
\hline 
Face area vector & $\mathbf{S}_{f}$ &  & Surface normal gradient & $\nabla_{S}$\tabularnewline
\hline 
Cell volume & $\mathcal{V}$ &  & Non-orthogonal correction & $\nabla_{\text{noc}}$\tabularnewline
\hline 
Label for a cell & $C$ &  & Interpolation from cells to faces & $\left\{ \right\} _{f}$\tabularnewline
\hline 
Label for a neighbour cell  & $N$ &  & Reconstruction of cell values from faces & $\left\{ \right\} _{C}$\tabularnewline
\hline 
Upwind label & $u$ &  & Hexagonal grid $n$ & HRn\tabularnewline
\hline 
Downwind label & $d$ &  & Cubed-sphere grid $n$ & Cn\tabularnewline
\hline 
Label for a face & $f$ &  & Horizontal resolution & $\Delta x$\tabularnewline
\hline 
Time-step & $\Delta t$ &  & Vertical resolution & $\Delta z$\tabularnewline
\hline 
High-order & HO &  & Temporal off-centring & $\alpha$\tabularnewline
\hline 
High-order correction & HOC &  & Implicit flag & $\beta$\tabularnewline
\hline 
Outer iteration counter & $k$ &  & High-order flux limiter & $\gamma$\tabularnewline
\hline 
\multicolumn{4}{|l|}{Note that $k$ is wavenumber in section \ref{subsec:NA}} & \tabularnewline
\hline 
\multicolumn{4}{|l|}{Vector from cell centre to cell centre over face $f$} & $\mathbf{d}_{f}$\tabularnewline
\hline 
\multicolumn{4}{|l|}{Diagonal matrix for the implicit conservation equation} & $A$\tabularnewline
\hline 
\multicolumn{4}{|l|}{Off-diagonal matrix for the implicit conservation equation} & $H$\tabularnewline
\hline 
\multicolumn{4}{|l|}{RHS of the matrix equation for the implicit conservation equation} & $R$ and $T$\tabularnewline
\hline 
\end{tabular}

\caption{Mathematical notation that is used in more than one section.\label{tab:notation}}
\end{table}

\section{Solution of the Fully Compressible Navier-Stokes Equations\label{appx:NSsolution}}

The potential temperature and momentum equations (\ref{eq:theta},\ref{eq:mom})
are first solved keeping the right hand side fixed and with the advection
and diffusion terms solved adaptively implicitly. Next, the advection
and diffusion terms are held fixed and (\ref{eq:theta},\ref{eq:mom})
are combined with the continuity equation (\ref{eq:cont}) and the
equation of state (\ref{eq:state}) to form a Helmholtz equation for
the perturbation Exner pressure, $\pi^{\prime}$. Prognostic variables,
$\mathbf{u}$, $\theta^{\prime}$ and $\rho$ are all at cell centres,
along with $\pi^{\prime}$. Mass fluxes through faces, $\phi$, are
calculated every time-step, with $\rho$ and $\mathbf{u}$ linearly
interpolated from cell centres onto faces in order to initialise $\phi$.

In order to describe the solution algorithm for each time-step, we
index the outer iterations with $k$, as in section \ref{subsec:ImExRK}.
At the beginning of each time-step, each variable at $k=0$ is set
to the old time level value, $n$. 

\subsection{Explicit Continuity update for consistency \label{subsec:continuityConsistency}}

The continuity equation is combined with the equation of state, potential
temperature and momentum equations to form a Helmholtz equation in
order to treat acoustic and gravity waves implicitly (section \ref{subsec:HelmholtzEqn}).
However, this process does not converge to machine precision, so the
resulting density and mass fluxes do not exactly solve the continuity
equation. This means that the flux form solutions of density weighted
variables (potential temperature and velocity) are not consistent
with continuity. In order to fix this, density is updated explicitly
using the fluxes from the Helmholtz equation (\ref{eq:Helmholtz})
before advecting mass weighted variables. Therefore, the first step
of the outer iteration, starting from $k=1$, is to update $\rho$
from the most recent flux, $\phi$.
\begin{equation}
\rho^{k}=\rho^{(n)}-\frac{\Delta t}{\mathcal{V}}\sum_{f}\left(1-\alpha\right)\phi^{(n)}-\frac{\Delta t}{\mathcal{V}}\sum_{f}\alpha\phi^{k-1}.\label{eq:expRho}
\end{equation}
This is an explicit update of a transport equation with a large Courant
number, which might be considered cause for concern. However the fluxes
used are implicit solutions of the continuity equation (section \ref{subsec:HelmholtzEqn}).
Therefore the solution of (\ref{eq:expRho}) is a small correction
and so does not lead to instability.

The mass fluxes, $\phi$, that are used in the solution of (\ref{eq:expRho})
for $\rho$ are the same as the mass fluxes and density that are used
in all advection-diffusion equations of the form (\ref{eq:advectDiffuse})
and solved in the form (\ref{eq:AandHpsi}). This gives consistency
with continuity and uniform $\psi$ fields are maintained.

\subsection{Adaptively Implicit Solution of the Potential Temperature Equation\label{subsec:solveTheta}}

Next the $\theta^{\prime}$ equation is solved to find $\theta^{\prime k}$
as described in section \ref{subsec:implicitAdvection}, with the
advection treated adaptively implicitly and the diffusion term treated
implicitly. The $\theta^{\prime}$ equation is put into the form (\ref{eq:AandHpsi})
with $K=\kappa$, with $A_{\theta^{\prime}}$, $H_{\theta^{\prime}}$
and $R_{\theta^{\prime}}$ as in (\ref{eq:Apsi}-\ref{eq:Rpsi}),
and with
\begin{eqnarray}
S_{\theta^{\prime}}^{(n)}=-\rho^{(n)}w^{(n)}\frac{d\overline{\theta}}{dz} &  & S_{\theta^{\prime}}^{k-1}=-\rho^{k}w^{k-1}\frac{d\overline{\theta}}{dz}.\label{eq:Stheta}
\end{eqnarray}
Note that $w$ from a previous iteration is used. This implies that
gravity waves are not treated implicitly in the solution of the $\theta^{\prime}$
equation. This comes in the solution of the momentum equation, described
in section \ref{subsec:solveMomentum}.

Solution of (\ref{eq:AandHpsi}) for $\theta^{\prime}$ gives $\theta^{\prime k}$.
From this we linearly interpolate to get $\theta_{f}^{\prime k}$
and add $\overline{\theta}$ to get $\theta^{k}$ and $\theta_{f}^{k}$.

\subsection{Adaptively Implicit Solution of the Momentum Equation\label{subsec:solveMomentum}}

Before the solution of the Helmholtz equation (section \ref{subsec:HelmholtzEqn}),
the momentum equation is solved with advection treated adaptively
implicitly, diffusion treated implicitly, gravity waves treated implicitly
and the pressure gradient held fixed from the previous iteration.
Without implicit gravity waves, the momentum equation fits into the
advection-diffusion equation form of (\ref{eq:advectDiffuse}) with
\begin{equation}
S_{\mathbf{u}}=-\rho\left\{ \frac{\left\{ \theta^{\prime}\right\} _{f}}{\overline{\theta}}\mathbf{g}\cdot\mathbf{S}_{f}+c_{p}\left\{ \theta\right\} _{f}\nabla_{S}\pi^{\prime}\right\} _{C}.\label{eq:SuExGravity}
\end{equation}
In order to treat gravity waves implicitly, part of the contribution
to $\theta_{f}^{\prime}$ from $w$ is removed so that it can be combined
with $w$ on the left hand side of the momentum equation. This leaves
a modified $\theta_{f}^{\prime}$ which is stored on faces and we
will call $\vartheta$
\begin{equation}
\vartheta=\overline{\theta}+\vartheta^{\prime}=\overline{\theta}+\theta_{f}^{\prime k}+\alpha\Delta tw_{f}^{k}\frac{\partial\overline{\theta}^{k}}{\partial z}.\label{eq:thetaWithoutw}
\end{equation}
$\theta_{f}^{\prime k}$ and $\theta_{f}^{k}$ are replaced by $\vartheta^{k}$
in (\ref{eq:SuExGravity}) using (\ref{eq:thetaWithoutw}) giving
\begin{equation}
S_{\mathbf{u}}=-\rho\alpha\Delta tw\frac{\partial\theta}{\partial z}\left(\frac{\mathbf{g}}{\overline{\theta}}+c_{p}\nabla\pi^{\prime}\right)-\rho\left\{ \frac{\vartheta^{\prime}}{\overline{\theta}}\mathbf{g}\cdot\mathbf{S}_{f}+c_{p}\vartheta\nabla_{S}\pi^{\prime}\right\} _{C}.\label{eq:Su}
\end{equation}
In order to treat gravity waves implicitly, vertical velocity, $w$,
and horizontal velocity, $\mathbf{v}$, are treated differently, with
the first term of (\ref{eq:Su}) being moved to $A_{w}$ for the $w$
equation. The equations for $\mathbf{v}$ and $w$ are of the form
(\ref{eq:AandHpsi}) with $R_{\mathbf{v}}$ and $R_{w}$ as in (\ref{eq:Rpsi})
$H_{\mathbf{v}}$ and $H_{w}$ as in (\ref{eq:Hpsi}), $A_{\mathbf{v}}$
as in (\ref{eq:Apsi}) and
\begin{eqnarray}
A_{w} & = & A_{\psi}-\rho^{k}\Delta t\frac{\partial\theta^{k}}{\partial z}\left\{ \alpha^{2}\left(\frac{\mathbf{g}\cdot\mathbf{S}_{k}}{\overline{\theta}_{f}}+c_{p}\nabla_{S}\pi^{\prime k-1}\right)\cdot\mathbf{k}\right\} _{C},\label{eq:Aw}\\
S_{\mathbf{v}}^{k-1} & = & -\rho^{k}\left\{ c_{p}\ensuremath{\vartheta}_{f}^{k}\nabla_{S}\pi^{\prime k-1}\right\} _{C},\label{eq:Sv}\\
S_{\mathbf{v}}^{(n)} & = & -\rho^{(n)}\left\{ c_{p}\theta_{f}^{(n)}\nabla_{S}\pi^{\prime(n)}\right\} _{C},\\
S_{w}^{k-1} & = & -\rho^{k}\left\{ \frac{\vartheta_{f}^{\prime k}}{\overline{\theta}}\mathbf{g}\cdot\mathbf{S}_{f}+c_{p}\theta_{f}^{k}\nabla_{S}\pi^{\prime k-1}\right\} _{C},\label{eq:Sw}\\
S_{w}^{(n)} & = & -\rho^{n}\left\{ \frac{\theta_{f}^{\prime(n)}}{\overline{\theta}}\mathbf{g}\cdot\mathbf{S}_{f}+c_{p}\theta_{f}^{(n)}\nabla_{S}\pi^{\prime(n)}\right\} _{C}.\label{eq:Swn}
\end{eqnarray}
Solution of (\ref{eq:AandHpsi}) for $\mathbf{v}$ and $w$ using
(\ref{eq:Aw}-\ref{eq:Swn}) gives us temporary values of $\mathbf{v}^{k}$
and $w^{k}$ which will be updated after the solution of the Helmholtz
equation (section \ref{subsec:HelmholtzEqn}).

\subsection{The Helmholtz Equation for Exner Pressure\label{subsec:HelmholtzEqn}}

Construction of the Helmholtz equation for $\pi^{\prime}$ involves
writing the mass flux, $\phi$, and the density, $\rho$, as a linear
function of $\pi^{\prime}$ and substituting both into the continuity
equation. 

To express $\phi$ as a linear function of $\pi^{\prime}$, we first
rewrite the velocities without contributions from the new time level
pressure gradient or gravity
\begin{eqnarray}
\tilde{\mathbf{v}}=\frac{H_{v}\mathbf{v}^{k}-R_{v}}{A_{v}} &  & \tilde{w}=\frac{H_{w}w^{k}-R_{w}}{A_{w}}.\label{eq:UHbyA}
\end{eqnarray}
To express $\phi$ in terms of $\tilde{\mathbf{v}}$ and $\tilde{w}$,
we interpolate onto the faces, add the contribution from the new time
level pressure gradient and gravity (note $\pi^{\prime k}$ not known
yet), multiply by density interpolated onto faces and take the dot
product with $\mathbf{S}_{f}$
\begin{equation}
\phi^{k}=\rho_{f}\left\{ \begin{array}{c}
\tilde{\mathbf{v}}\\
\tilde{w}
\end{array}\right\} _{f}\cdot\mathbf{S}_{f}-\alpha\Delta t\left\{ \begin{array}{cc}
\frac{\rho^{k}}{A_{v}} & 0\\
0 & \frac{\rho^{k}}{A_{w}}
\end{array}\right\} _{f}\left(\frac{\vartheta^{\prime}}{\overline{\theta}_{f}}\mathbf{g}+c_{p}\vartheta\nabla\pi^{\prime k}\right)\cdot\mathbf{S}_{f}.\label{eq:flux}
\end{equation}

To express $\rho$ as a linear function of $\pi^{\prime}$ we write
\begin{eqnarray}
\rho=\rho^{k}+\rho^{\prime}, &  & \pi^{\prime k}=\pi_{\rho}^{\prime}+\pi^{\prime\prime},
\end{eqnarray}
where $\rho^{\prime}$ is the small increment that updates $\rho^{k}$
to be consistent with $\pi^{\prime k}$ and $\pi_{\rho}^{\prime}$
satisfies the continuity equation with $\rho^{k}$
\begin{equation}
\rho^{k}=\frac{p_{r}}{R\theta}\left(\overline{\pi}+\pi_{\rho}^{\prime}\right)^{\frac{c_{v}}{R}},
\end{equation}
and $\pi^{\prime\prime}$ is the small increment that takes $\pi^{\prime}$
from $\pi_{\rho}^{\prime}$ to $\pi^{\prime k}$. Using the linearisation
\begin{equation}
\left(1+\frac{\pi^{\prime\prime}}{\overline{\pi}+\pi_{\rho}^{\prime}}\right)^{\frac{c_{v}}{R}}\approx1+\frac{c_{v}}{R}\frac{\pi^{\prime\prime}}{\overline{\pi}+\pi_{\rho}^{\prime}},
\end{equation}
we get the linearised equation of state
\begin{equation}
\rho=\rho^{k}\left(1+\frac{c_{v}}{R}\frac{\pi^{\prime\prime}}{\overline{\pi}+\pi_{\rho}^{\prime}}\right),\label{eq:linState}
\end{equation}
which can be substituted into the continuity equation to give a Helmholtz
equation for $\pi^{\prime k}=\pi_{\rho}^{\prime}+\pi^{\prime\prime}$
\begin{equation}
\frac{\rho^{k}\frac{c_{v}}{R}}{\overline{\pi}+\pi_{\rho}^{\prime}}\frac{\pi^{\prime k}-\pi_{\rho}^{\prime}}{\Delta t}+\frac{\rho^{k}-\rho^{(n)}}{\Delta t}+\nabla\cdot\phi^{k}=0,\label{eq:Helmholtz}
\end{equation}
where $\phi^{k}\left(\pi^{\prime k}\right)$ is from (\ref{eq:flux}).
(\ref{eq:Helmholtz}) is solved using the OpenFOAM multigrid solver
(GAMG) with a diagonal incomplete Cholesky (DIC) smoother until the
residual (the normalised volumetric mean absolute error, see WWKS23)
is below a tolerance. The tolerance is $10^{-6}$ for the final pressure
solve per time-step but capped at 0.01 times the initial residual
for other solves.

Next, $\pi^{\prime k}$ is used to back substitute to calculate $\phi^{k}$
from (\ref{eq:flux}) and to update the velocity
\begin{equation}
\mathbf{u}^{k}=\left(\begin{array}{c}
\tilde{\mathbf{v}}\\
\tilde{w}
\end{array}\right)-\Delta t\left(\begin{array}{cc}
\frac{\rho^{k}}{A_{v}} & 0\\
0 & \frac{\rho^{k}}{A_{w}}
\end{array}\right)\left\{ \alpha\left(\frac{\vartheta^{\prime}}{\overline{\theta}_{f}}\mathbf{g}\cdot\mathbf{S}_{f}+c_{p}\vartheta\nabla_{S}\pi^{\prime k}\right)\right\} _{C}.\label{eq:backU}
\end{equation}
All of the calculations in this subsection are repeated twice per
outer iteration with the calculation of $\tilde{\mathbf{v}}$ and
$\tilde{w}$ updated to use the most recent $\mathbf{v}^{k}$ and
$w^{k}$ for the second iteration.

\subsection{Algorithm Summary at each time-step}
\begin{itemize}
\item Calculate the Courant number in cells and on faces (\ref{eq:c},\ref{eq:cf})
and the Brunt--V\"{a}is\"{a}l\"{a} frequency, $N$(\ref{eq:N}).
From these calculate the stability parameters, $\alpha$, $\beta$
and $\gamma$ (table \ref{tab:parameterValues}).
\item Undertake two outer (Picard) iterations:
\begin{enumerate}
\item Update density, $\rho$, explicitly from mass fluxes, $\phi$, using
(\ref{eq:expRho}).
\item For $\theta^{\prime}$, calculate the explicit source, $S_{\theta^{\prime}}$
(\ref{eq:Stheta}), and the matrix diagonal and off-diagonal entries,
$A_{\theta}$ (\ref{eq:Apsi}), $H_{\theta}$ (\ref{eq:Hpsi}) and
the matrix source, $R_{\theta}$ (\ref{eq:Rpsi}) and $T_{\theta^{\prime}}$
(\ref{eq:Tpsi}).
\item Solve the matrix equation for $\theta^{\prime}$ (\ref{eq:AandHpsi})
using one CG iteration using a DILU preconditioner. In the final outer
iteration, this is repeated twice with explicit terms updated to improve
convergence.
\item Remove the part of the $\theta^{\prime}$ update related to gravity
waves using (\ref{eq:thetaWithoutw}) to calculate $\vartheta$ and
$\vartheta^{\prime}$.
\item For $\mathbf{u}$, calculate the explicit source terms, $S_{\mathbf{v}}$
and $S_{w}$ (\ref{eq:Sv},\ref{eq:Sw}) without the parts of $\theta^{\prime}$
that are treated implicitly for gravity waves. Calculate the matrix
diagonals, $A_{\mathbf{v}}$, $A_{w}$, with $A_{w}$ modified (\ref{eq:Aw})
to account for the implicit treatment of gravity waves. Calculate
the matrix off-diagonals, $H_{\mathbf{v}}$, $H_{w}$ and the matrix
source terms, $R_{\mathbf{v}}$, $T_{\mathbf{v}}$, $R_{w}$ and $T_{w}$
using (\ref{eq:Rpsi},\ref{eq:Tpsi}).
\item Solve the matrix equations for $u$, $v$ and $w$ (\ref{eq:AandHpsi})
using one CG iteration using a DILU preconditioner. These have pressure
gradients and positive buoyancy fixed from a previous iteration.
\item Two inner (Picard) iterations are used to solve the Helmholtz equation
for $\pi^{\prime}$ and to update $\mathbf{u}$:
\begin{enumerate}
\item Temporary values of the velocities, $\tilde{\mathbf{v}}$ and $\tilde{w}$,
are calculated without the pressure gradient or gravity using (\ref{eq:UHbyA}).
These are used to calculate the mass flux without the pressure gradient,
$\phi^{\prime}$, from (\ref{eq:flux}) but without the term involving
$\pi^{\prime}$.
\item $\phi^{\prime}$, and $\pi_{\rho}^{\prime}$ from the linearised equation
of state (\ref{eq:linState}), are substituted into the continuity
equation to get the Helmholtz equation (\ref{eq:Helmholtz}). This
is solved implicitly for Exner perturbation, $\pi^{\prime}$ using
the OpenFOAM multi-grid solver with an incomplete Cholesky smoother.
\item The back-substitution step updates $\phi$ from $\pi^{\prime}$ and
$\phi^{\prime}$ using (\ref{eq:flux}) and $\mathbf{u}$ from the
temporary velocities, $\tilde{\mathbf{v}}$ and $\tilde{w}$, using
(\ref{eq:backU}).
\end{enumerate}
\end{enumerate}
\item The stability parameters, $\alpha$, $\beta$ and $\gamma$, are only
calculated once per time-step. This means that the contributions from
the old time levels (terms including $1-\alpha$), only need to be
calculated once per time-step. This could potentially lead to stability
issues if the velocity or stable stratification were to dramatically
increase within one time-step. This does not occur in the test cases
in section \ref{sec:results}.
\end{itemize}

\section{Order of Convergence of Quasi-Cubic Advection Scheme\label{appx:cubicConvergence}}

In the deformational flow advection tests (section \ref{subsec:smoothTracer}),
the quasi-cubic advection scheme together with the adaptively implicit
time-stepping converges with second-order accuracy. However in a more
simplified setting, and with third-order time-stepping, third-order
convergence is achieved. 

The initial conditions consist of a double sinusoidal profile in a
unit square domain
\begin{equation}
\psi=\frac{1}{4}\left(1+\sin\left(2\pi x\right)\right)\left(1+\sin\left(2\pi y\right)\right)\ \left(x,y\right)\in[0,1)^{2},
\end{equation}
which is advected in a uniform grid by a constant velocity field,
$\left(1,1,0\right)$ for one time unit so that the initial conditions
return to their original location. The smoothness of the velocity
field and the initial conditions mean that initial conditions are
set as sampled values at cell centres and velocities are sampled at
face centres, rather than using third-order approximations to set
cell average and face average values. Grids of size $20\times20$,
$40\times40$, $80\times80$ and $160\times160$ are used with time-steps
of 0.02, 0.01, 0.005 and 0.0025 time units giving a Courant number
of 0.8. 

The advection uses third-order Runge-Kutta time-stepping
\begin{eqnarray}
\psi^{1} & = & \psi^{(n)}-\Delta t\nabla_{\text{HO}}\cdot\mathbf{u}\psi^{(n)},\nonumber \\
\psi^{2} & = & \psi^{(n)}-\frac{\Delta t}{4}\left\{ \nabla_{\text{HO}}\cdot\mathbf{u}\psi^{(n)}+\nabla_{\text{HO}}\cdot\mathbf{u}\psi^{1}\right\} ,\nonumber \\
\psi^{(n+1)} & = & \psi^{(n)}-\frac{\Delta t}{6}\left\{ \nabla_{\text{HO}}\cdot\mathbf{u}\psi^{(n)}+\nabla_{\text{HO}}\cdot\mathbf{u}\psi^{1}+4\nabla_{\text{HO}}\cdot\mathbf{u}\psi^{2}\right\} .
\end{eqnarray}

\begin{figure}
\noindent \begin{centering}
\includegraphics[width=0.7\textwidth]{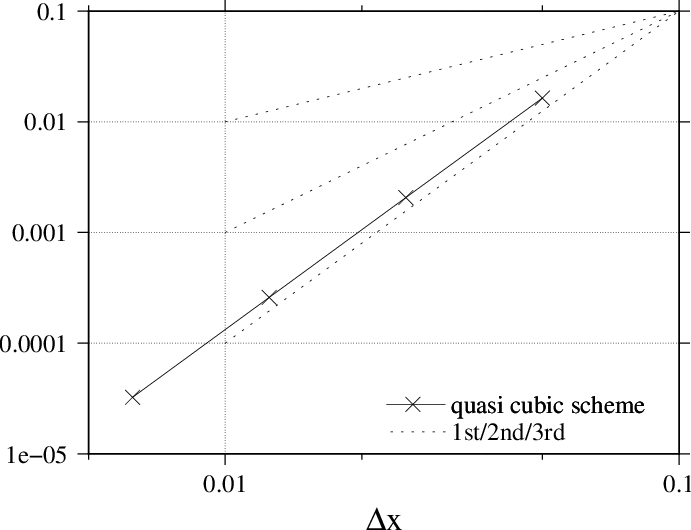}
\par\end{centering}
\caption{$\ell_{2}$ error norm after advecting a doubly sinusoidal profile
one revolution around a periodic, unit square by velocity field $\left(1,1,0\right)$
using quasi-cubic spatial discretisation and RK3 time-stepping. \label{fig:quasiCubicConvergence}}
\end{figure}

The $\ell_{2}$ error norm is the (normalised) root mean square error
and is shown as a function of resolution in figure \ref{fig:quasiCubicConvergence},
confirming the third-order convergence in this highly idealised setting.

\end{document}